\pdfoutput=1
\documentclass[aps,prd,amsmath,floats,floatfix,onecolumn,superscriptaddress,nofootinbib,showpacs]{revtex4-2}

\usepackage{comment}
\usepackage[T1]{fontenc}
\usepackage[utf8]{inputenc}
\usepackage{lmodern}
\usepackage{physics}
\usepackage[dvipsnames, usenames]{xcolor}
\definecolor{linkcolor}{rgb}{0.0,0.3,0.5}
\usepackage[hypertexnames=false, unicode, colorlinks=true, linkcolor=linkcolor,
citecolor=linkcolor, filecolor=linkcolor,urlcolor=linkcolor,
pdfusetitle]{hyperref}

\usepackage[all]{hypcap}
\usepackage{graphicx}
\usepackage{xspace}
\usepackage{amssymb}
\usepackage[normalem]{ulem} %for \sout
\usepackage{bm} % boldmath
\usepackage[small]{caption}
\usepackage{subcaption}
\usepackage{float}
\usepackage{titlesec}
\usepackage{multirow}

\usepackage{microtype}

\usepackage[english]{babel}
\graphicspath{%
  {figs/}%
}

\usepackage{tikz}
\usetikzlibrary{shapes,snakes}
\usetikzlibrary{arrows,intersections,patterns,decorations.markings,shapes.geometric}
\usetikzlibrary{calc,fadings,decorations.pathreplacing,positioning}
\usetikzlibrary{decorations.pathmorphing}
\tikzset{snake it/.style={decorate, decoration=snake}}

\usepackage{pgfplots}
\usepgfplotslibrary{colormaps}
\usetikzlibrary{intersections,patterns,pgfplots.fillbetween}

\tikzset{->-/.style={decoration={
  markings,
  mark=at position .5 with {\arrow{>}}},postaction={decorate}}
}
\tikzset{-<-/.style={decoration={
  markings,
  mark=at position .5 with {\arrow{<}}},postaction={decorate}}
}

\tikzset{%
  >=latex, % option for nice arrows
  inner sep=0pt,%
  outer sep=2pt,%
  mark coordinate/.style={inner sep=0pt,outer sep=0pt,minimum size=3pt,
    fill=black,circle}%
}

\DeclareMathAlphabet{\mathpzc}{OT1}{pzc}{m}{it}

\definecolor{darkred}{RGB}{175,0,0}
\definecolor{darkblue}{RGB}{14,0,185}
\definecolor{salmon}{RGB}{255,160,105}

\definecolor{darkblue}{RGB}{14,0,185}

\newcommand*{\bk}{\mathbf{k}}
\newcommand*{\bq}{\mathbf{q}}

\newcommand*{\bx}{\mathbf{x}}

\newcommand*{\ihMpc}{h\ \text{Mpc}^{-1}}

\newcommand{\beq}{\begin{equation}}
\newcommand{\eeq}{\end{equation}}

\def\q{{\boldsymbol{q}}}

\begin{document}

% --- Keep title OPTIONS (uncomment ONE of these when you decide) ---
% \title{Precision cosmological analysis using the LSS bootstrap}
% \title{Probing Nonlinear Structure Formation with the LSS Bootstrap}
% \title{Testing gravity and dark energy with the LSS bootstrap: a joint power spectrum and bispectrum analysis}
% \title{Beyond $\Lambda$CDM without a model: first data constraints on LSS bootstrap parameters from BOSS and PT Challenge}

% --- For now: keep ALL options visible in the compiled PDF (as in your draft) ---
\title{%
Probing nonlinear structure formation beyond $\Lambda$CDM with the LSS bootstrap: \\a joint power spectrum and bispectrum analysis
}
\begin{comment}
    \title{%
Precision cosmological analysis using the LSS bootstrap/Probing Nonlinear Structure Formation beyond $\Lambda$CDM with the LSS Bootstrap/%
Testing gravity with the LSS bootstrap: a joint power spectrum and bispectrum analysis/%
Beyond $\Lambda$CDM without a model: first data constraints on LSS bootstrap parameters from BOSS and PT Challenge Probing nonlinear structure formation beyond $\Lambda$CDM with the LSS bootstrap: a joint power spectrum and bispectrum analysis}
\end{comment}
\author{Giorgia Biselli}
\affiliation{Dipartimento di Scienze Matematiche, Fisiche e Informatiche, Universit\`a di Parma,
Parco Area delle Scienze, I-43124, Parma, Italy}
\affiliation{INFN Sezione Milano-Bicocca, Gruppo Collegato di Parma, I-43124, Parma, Italy}

\author{Marco Marinucci}
\affiliation{Institute for Theoretical Physics, ETH Zurich, 8093 Zurich, Switzerland}

\author{Guido D'Amico}
\affiliation{Dipartimento di Scienze Matematiche, Fisiche e Informatiche, Universit\`a di Parma,
Parco Area delle Scienze, I-43124, Parma, Italy}
\affiliation{INFN Sezione Milano-Bicocca, Gruppo Collegato di Parma, I-43124, Parma, Italy}

\author{Massimo Pietroni}
\affiliation{Dipartimento di Scienze Matematiche, Fisiche e Informatiche, Universit\`a di Parma,
Parco Area delle Scienze, I-43124, Parma, Italy}
\affiliation{INFN Sezione Milano-Bicocca, Gruppo Collegato di Parma, I-43124, Parma, Italy}

\date{\today} % keep empty, or use \date{\today}

\begin{abstract}
We present the first MCMC-derived constraints on the parameters of the Large Scale Structure (LSS)
bootstrap, a model-independent framework that captures deviations from $\Lambda$CDM using symmetry
arguments alone. Focusing on modifications to the linear growth rate and to the quadratic
perturbation-theory kernel -- quantified by the fractional parameters $\varepsilon_f$ and $\varepsilon_{d_{\gamma}}$,
respectively -- we carry out a joint analysis of the one-loop galaxy power spectrum and the tree-level
bispectrum multipoles within the EFTofLSS, employing the \texttt{PyBird} code extended to implement the
bootstrap parametrization. We apply this analysis pipeline to two datasets: the BOSS DR12 LRG sample
and the large-volume ``PT Challenge'' simulations. For BOSS, combining the power spectrum with the
bispectrum monopole yields $\sim 7\%$ constraints on $\varepsilon_f$ and $\sim 57\%$ constraints on
$\varepsilon_{d_{\gamma}}$. For the PT Challenge, whose survey volume is about 100 times larger, we reach
$\sim 1\%$ precision on $\varepsilon_f$ and $\sim 25\%$ on $\varepsilon_{d_{\gamma}}$, including the bispectrum quadrupole in the analysis. 
Our results underscore the complementary roles of $\varepsilon_f$ and $\varepsilon_{d_{\gamma}}$ in separating changes to the background
expansion from those affecting nonlinear structure formation, and they show that the LSS bootstrap
offers a competitive, model-agnostic method for probing physics beyond $\Lambda$CDM with existing and
upcoming galaxy surveys.
\end{abstract}

% Optional in REVTeX (you have showpacs enabled):
% \pacs{98.80.-k, 95.36.+x, 95.35.+d}
% \keywords{large-scale structure; EFTofLSS; bispectrum; modified gravity}

\maketitle

% --- Table of contents on the first page (as in your current draft) ---
\setcounter{tocdepth}{3} % include subsubsections in the ToC
\tableofcontents

% ... main text starts here ...
% \section{Introduction}

\section{Introduction}

Ongoing stage-IV galaxy surveys, including \textit{Euclid} \cite{Euclid:2024yrr} and DESI \cite{DESI:2016fyo}, are delivering an unprecedented volume of high-precision cosmological data on galaxy clustering. Maximizing the information extracted from these observations is crucial for subjecting the $\Lambda\text{CDM}$ paradigm to rigorous tests and for uncovering possible hints of physics beyond it. Theoretically, however, a wide variety of extensions to $\Lambda\text{CDM}$ has been put forward, encompassing modified gravity models and alternative dark energy frameworks. Each of these introduces additional parameters and distinct behaviors at both the background and perturbation levels. At the moment, there is no compelling theoretical argument for privileging any one model over the others. As a result, carrying out data analyses individually for every conceivable extension is computationally expensive and lacks strong theoretical motivation.
On the observational side, however, extensions of the standard paradigm have attracted increasing attention following the recent data releases from the DESI collaboration, which indicate a mild tension with the $\Lambda$CDM concordance model~\cite{DESI:2024mwx, DESI:2024jxi, DESI:2025zgx}.
The official analysis is often presented as a scenario in which dark energy evolves with time, but the result that the equation of state is $w < - 1$ conflicts with a single-field interpretation and therefore points to more complicated physics.
A discussion of additional possible extensions can be found in~\cite{DESI:2025fii}.

These challenges have motivated the construction of model-independent, symmetry-driven parameterizations, which make it possible to explore a broad range of cosmological models within a single, cohesive framework. This is the core principle of the \emph{Large Scale Structure (LSS) bootstrap}, originally introduced in \cite{DAmico:2021rdb} and later generalized in \cite{Marinucci:2024add, Peron:2025lgh} (see also \cite{Fujita:2020xtd} for earlier related work and~\cite{Ansari:2025nsf} for extensions). Instead of assuming a specific Lagrangian or fixed equations of motion, the LSS Bootstrap relies solely on the fundamental symmetries of our Universe -- such as the Equivalence Principle and rotational invariance -- to constrain the analytic structure of the perturbation theory kernels. In this way, it provides a general framework for analyzing large-scale structure data without being tied to specific cosmological scenarios.

In this work, we explore how effectively galaxy clustering measurements can constrain departures from the standard $\Lambda$CDM framework, expressed through a set of bootstrap parameters. A first step in this direction was undertaken in \cite{Amendola:2023awr} (see also \cite{Amendola:2022vte}), which provided the initial Fisher-matrix forecasts for these parameters for a Euclid-like survey.
Here, we perform a state-of-the-art combined analysis of the one-loop galaxy power spectrum and the tree-level bispectrum, both described within the Effective Field Theory of LSS (EFTofLSS) \cite{Baumann:2010tm, Carrasco:2012cv, perko2016biasedtracersredshiftspace}. Our numerical work relies on \texttt{PyBird} \cite{DAmico:2020kxu}, a fast Python code that computes redshift-space multipoles of biased tracers, which we extend to incorporate the bootstrap parameterization of the perturbation-theory (PT) kernels. The joint posterior distribution is then explored using Markov Chain Monte Carlo (MCMC) methods.

Several recent studies \cite{Philcox:2022frc, Ivanov:2023qzb, Bakx:2025pop} have highlighted the extra constraining power provided by bispectrum multipoles relative to analyses based solely on the power spectrum. For instance, \cite{Ivanov:2021kcd} reports roughly $5{-}15\%$ tighter bounds on cosmological parameters when the bispectrum monopole is incorporated, while \cite{DAmico:2022osl} demonstrates that including the one-loop monopole together with the tree-level quadrupole of the bispectrum yields an improvement of about $13{-}30\%$ compared with a power-spectrum-only analysis.
These works, formulated within the EFTofLSS, have been successfully confronted with both observational data and numerical simulations, and currently define the state-of-the-art in galaxy clustering studies~\cite{Chudaykin:2025aux, Chudaykin:2025lww, Chudaykin:2025vdh, Chudaykin:2026nls}. Nonetheless, they rest on the choice of a particular cosmological model. By contrast, the LSS Bootstrap offers a complementary, model-independent framework that enables an efficient exploration of a wide class of theories beyond $\Lambda\text{CDM}$.

We apply our pipeline  both on observational data from the BOSS survey \cite{reid2015sdssiiibaryonoscillationspectroscopic} and on mock galaxy catalogs constructed from the `PT Challenge' simulations \cite{Nishimichi:2020tvu}, which cover substantially larger volumes. In doing so, we can evaluate the range of constraints on new physics that can be derived from present-day data and from the measurements attainable by  a  futuristic survey.

The paper is organized as follows. In Sec.~\ref{sec:review_boot}, we review the Large-Scale Structure (LSS) Bootstrap and introduce the model-independent parameters used to detect possible departures from the $\Lambda\text{CDM}$ framework. 
%In Sec. \ref{sec:theoretical_ps_bisp}, we present the theoretical modeling of the galaxy power spectrum and bispectrum in redshift space. 
Sec.~\ref{sec:analysis} describes the datasets employed in the analysis, as well as the validation of the analysis pipeline, robustness tests against scale cuts and an investigation of the main degeneracies affecting the inference. The main results are presented in Sec.~\ref{sec:results}, and conclusions are summarized in Sec.~\ref{sec:conclusions}.

%%%%%%%%%%%%%%%%%%%%%%%%%%%%%%%

\section{The Large Scale Structure Bootstrap}
\label{sec:review_boot}
To establish the formalism for a model-independent study of extensions beyond $\Lambda$CDM, we start by briefly reviewing the \emph{Large Scale Structure (LSS) Bootstrap} (see \cite{DAmico:2021rdb, Marinucci:2024add, Peron:2025lgh} for more details). This framework offers a neat and systematic method to construct perturbative kernels based solely on symmetry arguments, thus ensuring broad applicability across a wide range of cosmological models. As in previous works, we assume Gaussian initial conditions and that the linear growth of perturbations is scale-independent.
Perturbations are defined, as usual, in term of the matter density contrast $\delta(\bx, a) \equiv \rho(\bx, a )/\rho_0 - 1$, where $\rho(\bx, a)$ is the matter energy density, $\rho_0$ its background value and $a$ is the scale factor, adopted here as our time variable. We further define the rescaled velocity divergence
\begin{equation}
    \theta(\bx, a) \equiv - \frac{\partial_iv^i(\bx, a)}{f(a)\mathcal{H}(a)},
\end{equation}
with $v^i(\bx, a)$ being the peculiar matter velocity and $\mathcal{H} = aH(a)$ the conformal Hubble rate. We have introduced the linear growth rate 
\begin{equation}
    f(a) = \frac{d\log{D(a)}}{d\log{a}}\,,
\end{equation}
where $D(a)$ is the usual linear growth factor, defined as $\delta_{\rm lin}(\bx, a) = D(a)/D(a_{\rm in})\,\delta_{\rm lin}(\bx, a_{\rm in})$.
We also consider the number density contrast of galaxies, 

\begin{equation}
    \delta_g(\bx, a) = \frac{n_g(\bx,a)}{\bar{n}_g} -1 \ ,
\end{equation}
where $\bar{n}_g$ is the galaxy mean number density.
These fields can be expressed as an expansion according to the standard perturbative ansatz \cite{Bernardeau:2001qr}

\begin{align}
    \delta(\bx, a) = \sum_{n=1}^{\infty}\delta^{(n)}(\bx, a) \ , \qquad 
    \theta(\bx, a) = \sum_{n=1}^{\infty}\theta^{(n)}(\bx, a) \ , \qquad \delta_g(\bx, a) = \sum_{n=1}^{\infty}\delta_g^{(n)}(\bx, a).
\end{align}

In Fourier space, these expansions can be expressed in terms of convolutions

\begin{align}
    \delta^{(n)}(\bk, a) & = \int\frac{d^3q_1}{(2\pi)^3} \ldots \int\frac{d^3q_n}{(2\pi)^3}(2\pi)^3\delta_D(\bk-\bq_{1\ldots n})F_n(\bq_1, \ldots, \bq_n; a)\delta^{(1)}(\bq_1; a)\ldots\delta^{(1)}(\bq_n; a) \ , \\
    \theta^{(n)}(\bk, a) & = \int\frac{d^3q_1}{(2\pi)^3} \ldots \int\frac{d^3q_n}{(2\pi)^3}(2\pi)^3\delta_D(\bk-\bq_{1\ldots n})G_n(\bq_1, \ldots, \bq_n; a)\delta^{(1)}(\bq_1; a)\ldots\delta^{(1)}(\bq_n; a) \ , \\
    \delta_g^{(n)}(\bk, a) & = \int\frac{d^3q_1}{(2\pi)^3} \ldots \int\frac{d^3q_n}{(2\pi)^3}(2\pi)^3\delta_D(\bk-\bq_{1\ldots n})K_n(\bq_1, \ldots, \bq_n; a)\delta^{(1)}(\bq_1; a)\ldots\delta^{(1)}(\bq_n; a)\,.
\end{align}

In the standard approach, the convolution kernels $F_n$, $G_n$, and $K_n$ are obtained by iteratively solving the continuity and Euler equations for a chosen cosmological model (their explicit expressions are given in Appendix \ref{app:kernel}). In the bootstrap approach, by contrast, the analytic form of these kernels is constrained by imposing symmetry principles, which leave a finite number of time- and cosmology-dependent undetermined functions.
For the matter and velocity kernels, we impose constraints that enforce rotational invariance, conservation of mass and momentum, and invariance under time-dependent spatial translations (the so-called extended Galilean invariance which, when combined with the adiabaticity  of long-wavelength modes, is equivalent to the Equivalence Principle (EP)). In addition, we require perturbativity, namely that the kernels are built from products of external momentum invariants, with the only permissible poles taking the form $\bq_i/q_i^2$ (for more details, see \cite{Marinucci:2024add}). 

As a result, at second order, the symmetric matter and velocity kernels take the form
\begin{align}
    F_2(\bq_1, \bq_2; a) & = \beta(\bq_1, \bq_2) + \frac{a_{\gamma}^{(2)}(a)}{2}\gamma(\bq_1, \bq_2) \ , \label{F2} \\
    G_2(\bq_1, \bq_2; a) & = \beta(\bq_1, \bq_2) + \frac{d_{\gamma}^{(2)}(a)}{2}\gamma(\bq_1, \bq_2)
    \label{G2} \ ,
\end{align}
where the mode-coupling functions are defined as
\begin{align}
    \beta(\bq_1, \bq_2) & = \frac{|\bq_1+\bq_2|^2\bq_1\cdot \bq_2}{2q_1^2q_2^2} \ , \\
    \gamma(\bq_1, \bq_2) & = 1 - \frac{(\bq_1\cdot \bq_2)^2}{q_1^2q_2^2} \ . 
\end{align}
The coefficients of $\beta(\bq_1,\bq_2)$ are fixed to one by the EP, whereas the undetermined, time-dependent coefficients $a_{\gamma}^{(2)}$ and $d_{\gamma}^{(2)}$ that appear in \eqref{F2} and \eqref{G2} carry information about the underlying cosmological model.  Once a specific model is chosen, these functions can be derived from their equations of motion; for example, in $\Lambda$CDM they are given by \cite{DAmico:2021rdb}
\begin{align}
    \frac{d \, a_{\gamma}^{(2)}(a)}{d \ln{a}} &= f \Big(2 - 2a_{\gamma}^{(2)} + d_{\gamma}^{(2)} \Big) \ , \\
    \frac{d \, d_{\gamma}^{(2)}(a)}{d \ln{a}} &= f \Bigg(\frac{3}{2}\frac{\Omega_m}{f^2} \Big(a_{\gamma}^{(2)} - d_{\gamma}^{(2)} \Big) - d_{\gamma}^{(2)} \Bigg) \ .
    \label{eq:bootEoM}
\end{align}
The PT kernels for biased tracers, the $K_n$'s, can be obtained in an analogous way to the matter case, with the crucial difference that mass and momentum conservation are not enforced. Because of this, the $K_n$ kernels involve a larger set of independent coefficients than the corresponding matter and velocity kernels at the same perturbative order. Moreover, since the residual symmetries (EP and rotational invariance) coincide with those used in constructing the bias expansion \cite{Desjacques:2016bnm}, the bootstrap coefficients for biased tracers are completely degenerate with the bias parameters themselves. They therefore cannot be used to recover any cosmological information, i.e., information that is independent of the tracer. As a result, within the bootstrap framework, cosmological constraints arise solely from the velocity sector through redshift-space distortions (RSD), via the linear growth rate $f$, and the quadratic perturbative parameter $d_{\gamma}^{(2)}$.

\noindent In general, more bootstrap terms appear at third order: specifically $d_{\gamma a}^{(3)}$, $d_{\gamma b}^{(3)}$ and $a_{\gamma}^{(2)}$, see appendix~\ref{app:kernel} for their definitions. Given a specific model, one would generally expect (correlated) deviations in all these coefficients, not just $f$ and $d_{\gamma}^{(2)}$.
Within the analysis performed in this work, power spectrum at one-loop and bispectrum at tree-level, these terms enter only through the $P_{13}$ term and are expected to be very degenerate with the third order bias parameter. Moreover, our combination of observables is particularly insensitive to such third order bias and bootstrap terms, while it is well suited to constrain second order terms, such as quadratic and tidal bias~\cite{DAmico:2022osl, Ivanov:2021kcd, Philcox:2021kcw, Philcox:2022frc, Ivanov:2023qzb}. For these reasons, we will keep all the third order bootstrap coefficient fixed to their $\Lambda$CDM prediction and limit our analysis only to deviations from $\Lambda$CDM to linear and quadratic order in the perturbative expansion, $f(a)$ and $d_{\gamma}^{(2)}(a)$. We introduce the fractional deviation parameters,
\begin{equation}
    \varepsilon_f  \equiv \frac{f}{f^{\Lambda\text{CDM}}} - 1\,, \qquad\qquad \varepsilon_{d_{\gamma}}  \equiv \frac{d_{\gamma}^{(2)}}{d_{\gamma}^{(2),\,\Lambda\text{CDM}}} -1 ,
    \label{eq:eps}
\end{equation}
where the dependence of the $\Lambda$CDM quantities in the denominators on $\Omega_m$, given by Eq.~\eqref{eq:bootEoM}, is taken into account. Any detected non-zero value of $\varepsilon_f$ or $\varepsilon_{d_{\gamma}}$ would therefore signal new physics beyond $\Lambda$CDM, which cannot be reabsorbed by merely adjusting $\Omega_m$. By including $d_\gamma^{(2)}$, we probe deviations from $\Lambda$CDM in the nonlinear regime, thereby extending the analysis past the linear parameter $f$ and enhancing our ability to distinguish among different beyond-$\Lambda$CDM scenarios. As emphasized in previous work~\cite{Amendola:2023awr}, extensions of $\Lambda$CDM within modified gravity or evolving dark energy scenarios can lead to distinct signatures: models that primarily modify the background expansion, such as those involving non-standard dark energy evolution, mostly affect the time dependence of growth function, resulting in a non-zero $\varepsilon_f$, while leaving the nonlinear evolution of perturbations largely unchanged, i.e. $\varepsilon_{d_\gamma} \simeq 0$. In contrast, modified gravity models typically alter both the background dynamics and the growth of nonlinear structure, leading to a characteristic signal with $\varepsilon_{d_\gamma} \neq 0$. This framework therefore provides a model-independent way to disentangle modifications to the background evolution, encoded in $\varepsilon_f$, from those affecting nonlinear growth, described by $\varepsilon_{d_\gamma}$, thereby clarifying the theoretical requirements that viable extensions of $\Lambda$CDM must satisfy.

To connect the model-independent deviations $\varepsilon_f$ and $\varepsilon_{d_{\gamma}}$ with observables, the galaxy power spectrum and bispectrum are modeled using the EFTofLSS \cite{Baumann:2010tm, Carrasco:2012cv}, following the implementation in \texttt{PyBird} \cite{DAmico:2020kxu}, modified to include the bootstrap framework and the time-dependence of the kernels. The explicit form for the observables is given in Appendix \ref{sec:theoretical_ps_bisp}.

We apply the Alcock-Paczynski (AP) effect \cite{Alcock:1979mp} as implemented in \texttt{PyBird} to all the analyses (see \cite{DAmico:2022osl} for details).

%%%%%%%%%%%%%%%%%%%%%%%%%%%%%%%%%%%%%%%%%%

\section{Datasets and tests}
\label{sec:analysis}
In this section, we describe the datasets and methodology employed to constrain the model-independent parameters $\varepsilon_f$ and $\varepsilon_{d_{\gamma}}$, using both observational data and numerical simulations.

\subsection{BOSS analysis} 
\label{sec:BOSS_analysis}
We analyze the redshift-space monopole, quadrupole and hexadecapole of the power spectrum, together with the monopole of the bispectrum of the SDSS III BOSS DR12 luminous red galaxies (LRG) sample at redshifts $z=0.32$ (LOWZ) and $z=0.57$ (CMASS) \cite{Alam_2017, reid2015sdssiiibaryonoscillationspectroscopic}, using data from both the Northern (NGC) and the Southern (SGC) Galactic Caps.

Computing the power spectrum at the central value of each wavenumber bin, instead of averaging it over the full bin width, has a negligible effect for BOSS. We therefore disregard this difference for the BOSS data, in contrast to the PT Challenge data (see below). For the bispectrum, we adopt the same binning strategy as in \cite{DAmico:2022osl}. 
We estimate the covariance matrix, following~\cite{DAmico:2022osl}, for the power spectrum, the bispectrum and their cross correlation using Patchy mock catalogs~\cite{Kitaura_2016}. This catalog consists of 2048 simulated realizations built using the halo occupation distribution (HOD) algorithm in order to match the BOSS galaxy clustering and the different selection functions for each redshift bin. The covariance matrices are estimated using the usual expression
\begin{equation}
    \text{Cov}\big(O_{\ell}(m_i)O_{\ell'}(m_j)\big) = \frac{1}{N_{\rm sims} - 1}\sum_{n=1}^{N_{\rm sims}} \Big[O_{\ell}^{(n)}(m_i) - \bar{O}_{\ell}(m_i)\Big]\Big[O_{\ell'}^{(n)}(m_j) - \bar{O}_{\ell'}(m_j)\Big]\,,
\end{equation}
where $O = \{P, B\}$ are the observables for which we estimate the covariance, $m_i = \{k_i, t_i\}$, respectively the $i$-th $k$-bin and the $i$-th triangular bin. $N_{\rm sims}$ is the total number of simulations, the superscript $(n)$ indicates the measured multipole of the $n$-th realization and the average is defined as
\begin{equation}
    \bar{O}_{\ell}(m) \equiv\frac{1}{N_{\rm sims}}\sum_{n=1}^{N_{\rm sims}} O_{\ell}^{(n)}(k)\,.
\end{equation}

%%%%%%%%%%%%%%%%%%%%%%%%%%%%%%%%%%%%%%%%%%

\subsection{PT Challenge simulations}
\label{subsec:PTch}
We also analyze the redshift-space power spectrum and bispectrum monopole and quadrupole of the `PT Challenge' simulations \cite{Nishimichi:2020tvu}. 
%\GB{We do not include the power spectrum hexadecapole to the analysis, as its signal-to-noise ratio is small and, as shown in App.~\ref{app:hexadecapole}, it does not significantly improve the results.}
They consist of ten realizations in periodic comoving boxes, with a combined volume of $566 \, (h^{-1}\,\text{Gpc})^3$. This corresponds to a volume roughly two orders of magnitude larger than that of the BOSS catalogs and about $O(10)$ times greater than that probed by Stage IV surveys. The simulations adopt a flat $\Lambda$CDM cosmology, allowing us to carry out null tests of the bootstrap parameters \eqref{eq:eps}. This enables us to evaluate both the constraining power of an ideal, extremely large-volume survey that is free from real-world systematic effects, and the robustness of our methodology.

Dark matter halos are populated using the halo occupation distribution (HOD) technique, matched to reproduce the observed clustering properties of the BOSS samples, in particular the monopole moment of the power spectrum. In this work, we focus on a snapshot at redshift $z = 0.61$, that matches the properties of the ``high-z'' BOSS data sample. 

In contrast to the BOSS analysis, the substantially  smaller uncertainties of the PT Challenge simulations make the effect of binning of the power spectrum non-negligible.
We therefore account for this effect as prescribed by  the \textit{West Coast team} in \cite{Nishimichi:2020tvu}. As for the bispectrum, we consider triangles whose sides start from  the fundamental frequency $k_f = 0.005 \,\ihMpc$, and are binned with a linear spacing of width $\Delta k = 2 k_f $.

The power spectrum covariance matrix is supplied with the data and is computed under the assumption of Gaussianity, following the methodology of \cite{Nishimichi:2020tvu}, separately for each of the ten realizations. For the analysis of the total volume spanned by the ten boxes, we combine these ten individual covariance matrices into a single covariance via a weighted average.

The covariance matrix for the bispectrum is estimated as described in App. \ref{app:theoretical_cov}.
We neglect the cross-covariance between the power spectrum and bispectrum, as it has been shown to be negligible at the scale cuts adopted in this work (see, e.g., \cite{Wadekar:2020hax, Ivanov:2021kcd, Rizzo:2022lmh, Philcox:2022frc, Ivanov:2023qzb}). However, we include the correlations between different bispectrum multipoles, as done for the power spectrum multipoles $P_{\ell}.$

%%%%%%%%%%%%%%%%%%%%%%%%%%%%%%%%%%%%%%%%%%

\subsection{Parameters and priors}
\label{sec:param_priors}

We perform MCMC analyses using the Metropolis–Hastings algorithm as implemented in \texttt{MontePython 3} \cite{brinckmann2018montepython3boostedmcmc, Audren:2012wb}, numerically sampling the parameter set
\begin{equation}
    \{\omega_b, \omega_{cdm}, h, \ln{(10^{10}A_s)}, \varepsilon_f, \varepsilon_{d_{\gamma}}, b_1, b_2, b_4 \}.
\end{equation}
We follow the \texttt{PyBird} \cite{DAmico:2020kxu} convention for the quadratic bias parameters $b_2$ and $b_4$; see also Eq.~\eqref{eq:b24K} for their mapping to alternative bases.
We impose flat priors on all parameters except for the baryonic density $\omega_b$ (see below) and the primordial scalar amplitude $A_s$. Since $A_s$ exhibits strong degeneracies (see Sect.~\ref{sec:degeneracies}), we consider three different treatments:
\begin{enumerate}
\item we leave it free (only for the BOSS analysis);
    \item we fix $\ln{(10^{10}A_s)} = 3.0448$, the \textit{Planck} best-fit value \cite{Planck2018}, for the BOSS analysis, and to the true value used in the simulations for the PT Challenge;
    \item we instead adopt a Gaussian prior on $A_s$ with a width equal to $3\sigma$ of the \textit{Planck} uncertainty \cite{Planck2018}, following refs.~\cite{Tsedrik:2022cri, Carrilho:2022mon}.
\end{enumerate}

Given that \textit{Planck}'s uncertainty on $A_s$ is extremely small, fixing $A_s$ or applying a $3\sigma$ Gaussian prior leads to nearly identical constraints, as demonstrated in App.~\ref{app:As_comparison}.

The BOSS data are split into two separate redshift samples, and the model-independent parameters $\varepsilon_f$ and $\varepsilon_{d_{\gamma}}$ are in general time-dependent. We therefore treat them as independent in each redshift bin, obtaining two sets of bootstrap parameters per redshift:
\begin{equation}
    \{\varepsilon_f^l, \varepsilon_{d_{\gamma}}^l, \varepsilon_f^h, \varepsilon_{d_{\gamma}}^h \},
\end{equation}
where the superscripts $l$ and $h$ denote the LOWZ and CMASS samples, respectively. The bias parameters $(b_1, b_2, b_4)$ are treated as independent for each redshift bin and sky region (NGC and SGC), resulting in four sets of bias parameters.
Conversely, the cosmological parameters are assumed to be shared across both bins and sky regions.

Adopting the notation of \cite{DAmico:2022ukl, DAmico:2022gki}, we perform an analytic marginalization over the following bias and EFT parameters that appear in the theoretical model (see Appendix~\ref{sec:theoretical_ps_bisp} for details):
\begin{equation}
    \{c_{\text{ct},1}, c_{\text{ct},2}, c_{\text{ct},3}, c_{\text{ct},4} , c_{\epsilon,0},c_{\epsilon,1},c_{\epsilon,\text{quad}}, c_{\epsilon,3}, c_{\epsilon,4}, \tilde{b}_3, \tilde{b}_8 \}
\end{equation}
where we have defined the following combination of parameters:
\begin{align}
    c_{\epsilon,\text{quad}} = \frac{2}{3} f c_{\epsilon,2} \ ,  \qquad
    c_{\text{ct},1} = - c_{\text{ct}} \ , \qquad
    c_{r,1} = c_{\text{ct},2} f- \frac{1}{2}f^2 c_{\text{ct},4} \ , \qquad
    c_{r,2}  = - \frac{1}{2} c_{\text{ct},3} f^2 \ , \qquad  b_3 = \tilde{b}_3 + 15 \ \tilde{b}_8.
\end{align}
The mapping between the parameters adopted in this work and those used in \texttt{PyBird} can be found in Appendix D4 of \cite{DAmico:2022ukl}.
In our BOSS data analysis we fix the scalar spectral index to the \textit{Planck} best-fit value, $n_s = 0.96605$,  since the data are not very sensitive to it. We apply a Gaussian prior on the baryon density, $\omega_b = 0.022445 \pm 0.00038$, as inferred from Big Bang Nucleosynthesis \cite{Planck2018}.
For the PT Challenge analysis, we hold the spectral index fixed at the fiducial value used in the simulations, $n_s = 0.9649$\footnote{See the \href{https://www2.yukawa.kyoto-u.ac.jp/~takahiro.nishimichi/data/PTchallenge/}{PT Challenge website} for details}, and apply a Gaussian prior on the baryon density $\omega_b$, again centered on the fiducial value adopted in the simulations with the BBN width, 0.00038.

For the strong-coupling scale $k_M$ (see Appendix \ref{sec:theoretical_ps_bisp}) we adopt the value $k_M = 0.7\, \ihMpc$.
The galaxy number density is set to $\bar{n}_g = 4 \cdot 10^{-4} (h^{-1}\,\text{{Mpc}})^3$ for BOSS and $\bar{n}_g = 3 \cdot 10^{-4} (h^{-1}\,\text{{Mpc}})^3$ for the PT Challenge simulations.
We adopt flat priors on the bootstrap parameters  
\begin{align}
    \varepsilon_f &\in [-2,2]\,, \nonumber \\
    \varepsilon_{d_{\gamma}} &\in [-2,2],
\end{align}
and Gaussian priors on all bias and EFT parameters, as listed in Table \ref{tab:priors}.
\begin{table}[t]
\centering
\begin{tabular}{|c|c|c|}
\hline
Parameter & Prior (PT Challenge) & Prior (BOSS) \\
\hline
$b_1$ & $\text{Lognormal}(0.8,0.8)$ & - \\
$b_2$     & $\mathcal{N}(0,2)$ & - \\
$b_4$     & $\mathcal{N}(0,2)$ & - \\
$\tilde{b}_3$       & $\mathcal{N}(0,2)$ & - \\
$\tilde{b}_8$       & $\mathcal{N}(0,2)$ & - \\
$c_{\text{ct},1}$ & $\mathcal{N}(0,4)$ & - \\
$c_{\text{ct},2}$ & $\mathcal{N}(0,4)$ & - \\
$c_{\text{ct},3}$ & $\mathcal{N}(0,4)$ & - \\
$c_{\text{ct},4}$ & $\mathcal{N}(0,4)$ & - \\
$c_{\epsilon,0}$ & $\mathcal{N}(1,0.2)$ & $\mathcal{N}(0,0.2)$ \\
$c_{\epsilon,1}$ & $\mathcal{N}(0,4)$ & - \\
$c_{\epsilon,\text{quad}}$ & $\mathcal{N}(0,2)$ & - \\
$c_{\epsilon,3}$ & $\mathcal{N}(1,0.4)$ & $\mathcal{N}(0,0.4)$ \\
$c_{\epsilon,4}$ & $\mathcal{N}(1,1)$ & $\mathcal{N}(0,1)$ \\
\hline
\end{tabular}
\caption{List of priors adopted on bias, counterterms, and stochastic parameters. $\mathcal{N}(\mu, \sigma)$ represents a normal distribution with mean $\mu$ and standard deviation $\sigma$. A log-normal prior is adopted for $b_1$, implemented as a Gaussian prior on $\ln{b_1}$. A dash (-) indicates that the corresponding parameter shares the same prior as in the PT Challenge case.}
\label{tab:priors}
\end{table}

Overall, our choice of priors largely mirrors those adopted in earlier \texttt{PyBird} analyses \cite{Nishimichi:2020tvu, Piga:2022mge, DAmico:2022gki, DAmico:2022osl}. 
The only departure concerns the stochastic parameters in the PT Challenge analysis. 
In the PT Challenge data provided to us, the shot noise has been removed from the power spectrum but not from the bispectrum. To maintain consistency between the two, we reintroduced the shot-noise term into the power spectrum monopole. Specifically, we estimated the mean number density in each simulation box via
\begin{equation}
     \frac{1}{\bar{n}} = \frac{L^3}{N_{\text{gal}}}
\end{equation}
with $L = 3840 \,h^{-1} \,\text{Mpc}$ denoting the box size and $N_{\text{gal}}$ the number of mock galaxies in the box. 
We then averaged $1/\bar{n}$ over the 10 boxes and added this constant shot-noise correction to the power spectrum monopole in every k-bin. Since we explicitly restore the $1/\bar{n}$ contribution to the power spectrum monopole, we shift the prior on the associated stochastic nuisance parameters $c_{\epsilon,0}, c_{\epsilon,3}, c_{\epsilon,4}$, as defined in Appendix \ref{sec:theoretical_ps_bisp}, to be centered on $1$ rather than $0$.

We declare the convergence of the MCMC's when the Gelman-Rubin $R-1$ value \cite{Gelman:1992zz} is $\leq 0.03$ for all sampled parameters. 
All our plots are generated using the $\texttt{GetDist}$ package \cite{Lewis_2025}.

%%%%%%%%%%%%%%%%%%%%%%%%%%%%%%%%%%%%%%%%%%

\subsection{Scale cuts}
\label{sec:scale_cuts}

We do not perform scale-cut tests for the BOSS data, since we adopt the same maximum wavenumbers $k_{\text{max}}$ that were already validated in previous \texttt{PyBird}-based analyses of the BOSS DR12 LRG sample \cite{DAmico:2019fhj, DAmico:2022osl, DAmico:2022ukl}. Concretely, we use
$k_{\text{max},P_\ell}=0.20\,h\,\text{Mpc}^{-1}$ for LOWZ, $k_{\text{max},P_\ell}=0.23\,h\,\text{Mpc}^{-1}$ for CMASS, and 
$k_{\text{max},B_0}=0.08\,h\,\text{Mpc}^{-1}$ for the bispectrum monopole in both samples.

For the PT Challenge simulations, following the arguments in \cite{Nishimichi:2020tvu}, we analyze the power spectrum multipoles up to $k_{\text{max}} = 0.14\, \ihMpc$. In contrast, for the bispectrum multipoles we explicitly test different values of $k_{\text{max}}$, denoted $k_{\text{max}, B_0}$ and $k_{\text{max}, B_2}$ for the monopole and quadrupole, respectively. These tests are carried out using the parameter set and priors specified in Sec.~\ref{sec:param_priors}, while fixing the primordial scalar amplitude $A_s$ to the true value adopted in the simulations.

\begin{figure}[H]
  \centering
  \begin{subfigure}{0.75\textwidth}
    \includegraphics[width=\linewidth]{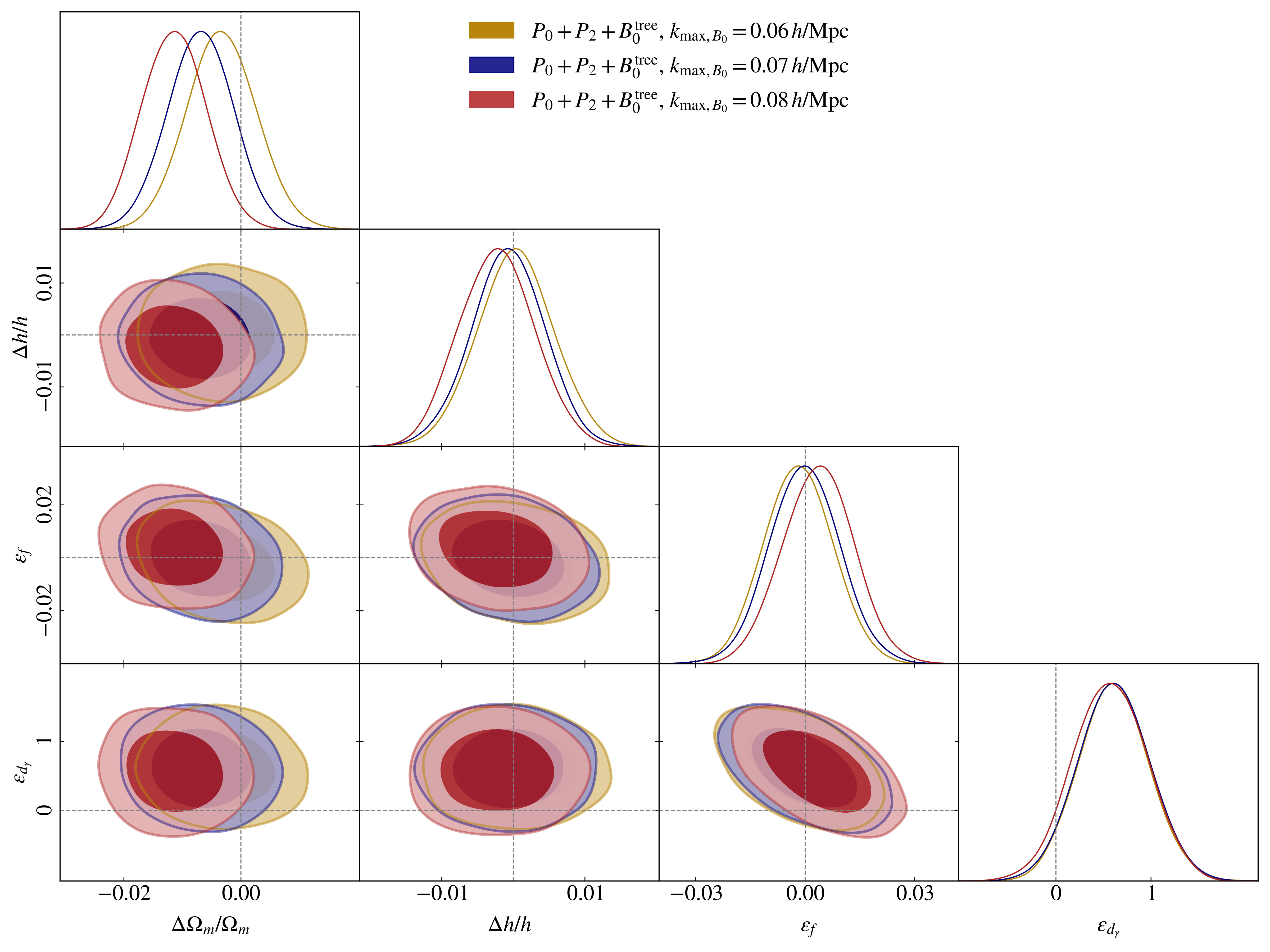}
    \caption{Posteriors as a function of $k_{\text{max},B_0}$ for the PT Challenge simulations.}
    \label{fig:k_max_mon_cov_th}
  \end{subfigure}
  \hfill
  \begin{subfigure}{0.75\textwidth}
    \includegraphics[width=\linewidth]{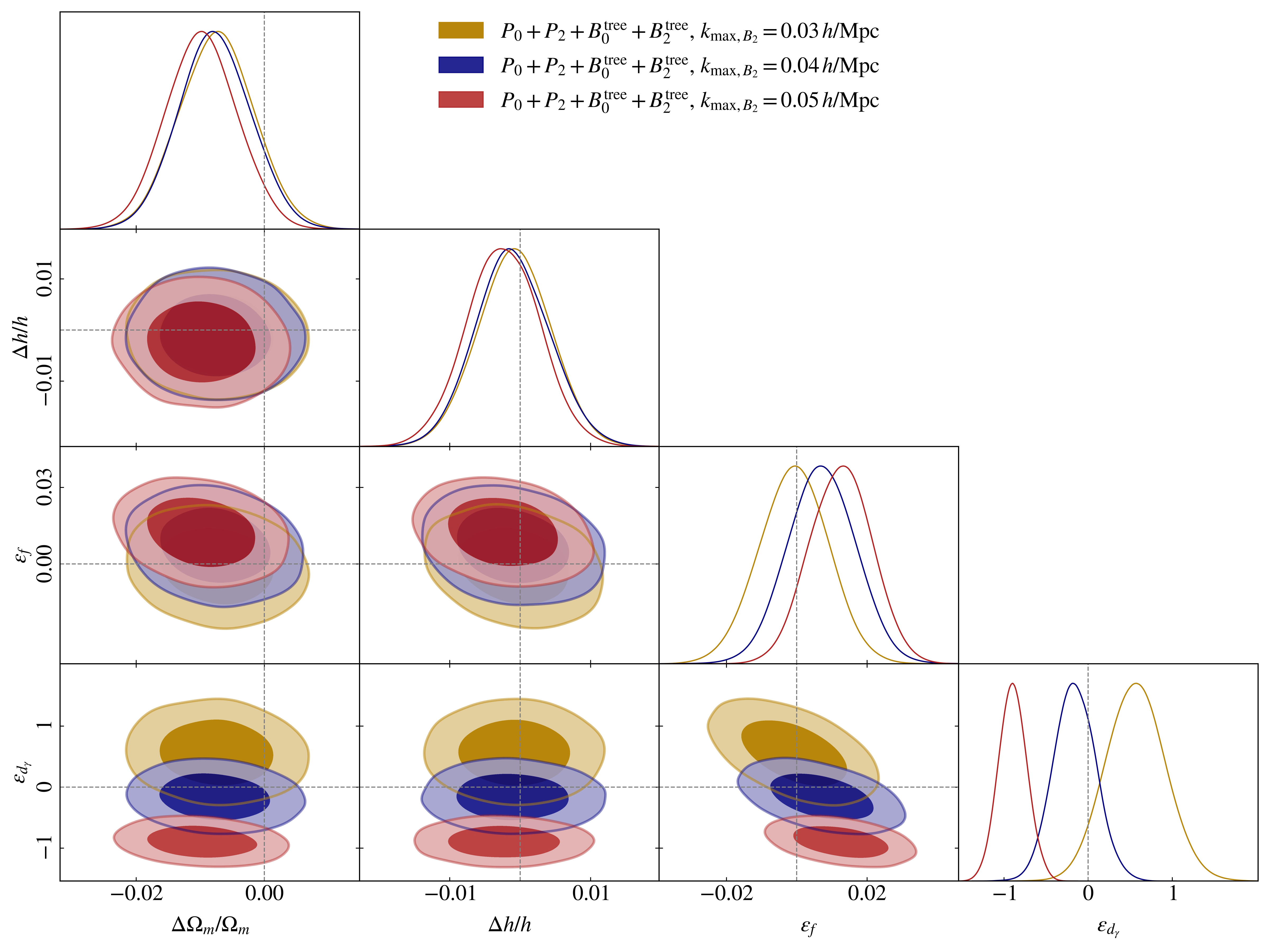}
    \caption{Posteriors as a function of $k_{\text{max},B_2}$ for the PT Challenge simulations. In all cases, we fix $k_{\text{max},B_0} = 0.07 \, \ihMpc$.}  
    \label{fig:k_max_quad_cov_th}
  \end{subfigure}
 \caption{Impact of the choice of $k_{\text{max}}$ on the posterior distributions for the PT Challenge simulations for the bispectrum monopole $B_0$ (top panel) and quadrupole $B_2$ (bottom panel). Both analyses are performed with $n_s$ and $A_s$ fixed to the true values of the simulations.}
  \label{fig:kmax_cov_th}
\end{figure}

Figure \ref{fig:k_max_mon_cov_th} shows the effect of raising $k_{\text{max}, B_0}$ from $0.06$ to $0.08 \,\ihMpc$. Increasing this cut leads to shifts in the posterior means of the $\varepsilon$-parameters, but these remain well within their corresponding $1\sigma$ statistical errors. In contrast, the inferred value of $\Omega_m$ shifts by more than $1\sigma$ once $k_{\text{max}, B_0}$ exceeds $0.07 \,\ihMpc$, signaling the growing importance of 1-loop contributions, as discussed below.

\begin{figure}[H]
  \centering
  \begin{subfigure}{0.49\textwidth}
    \includegraphics[width=\linewidth]{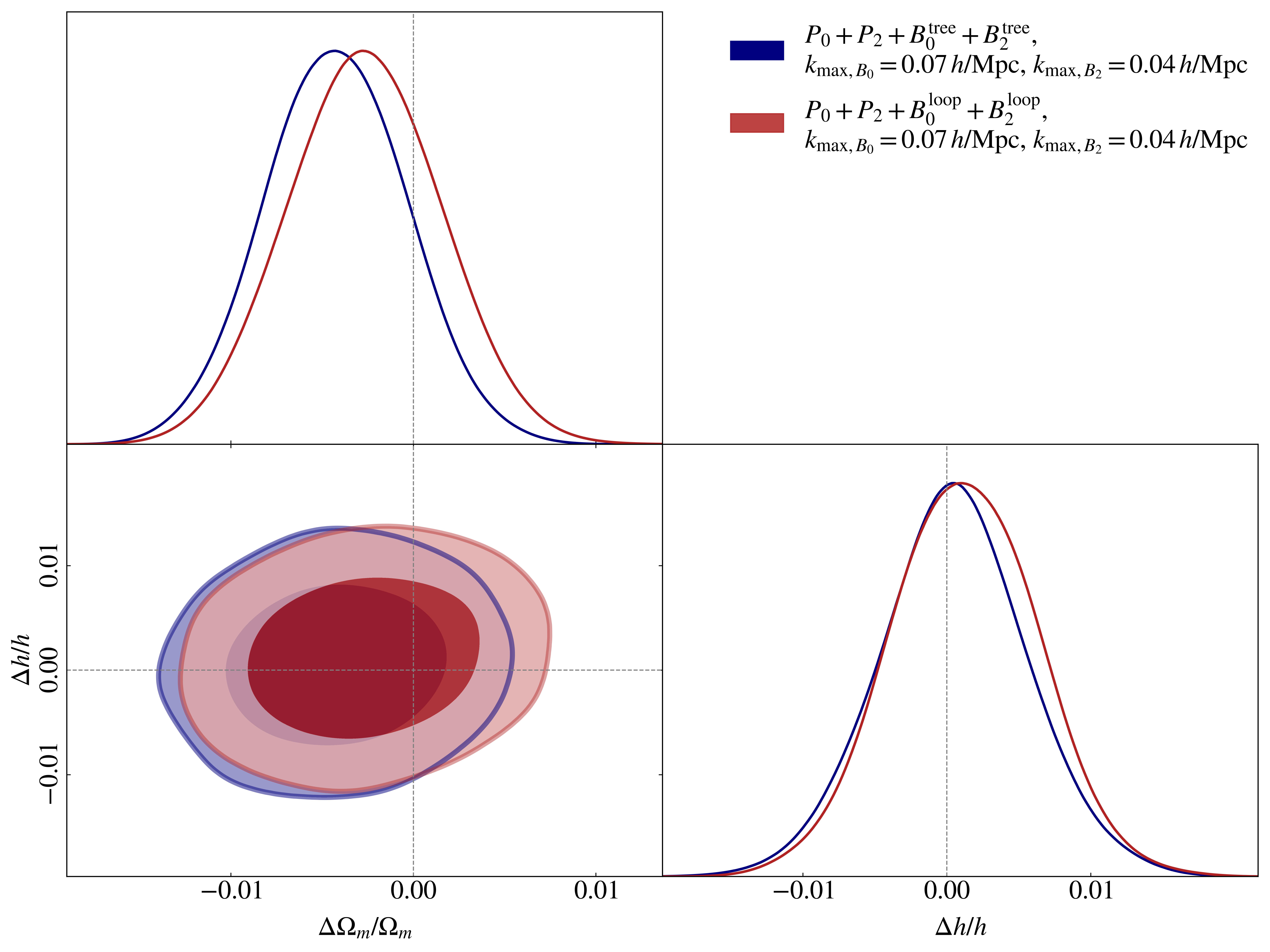}
    \caption{}
    \label{fig:quad_loop_004}
  \end{subfigure}
  \hfill
  \begin{subfigure}{0.49\textwidth}
    \includegraphics[width=\linewidth]{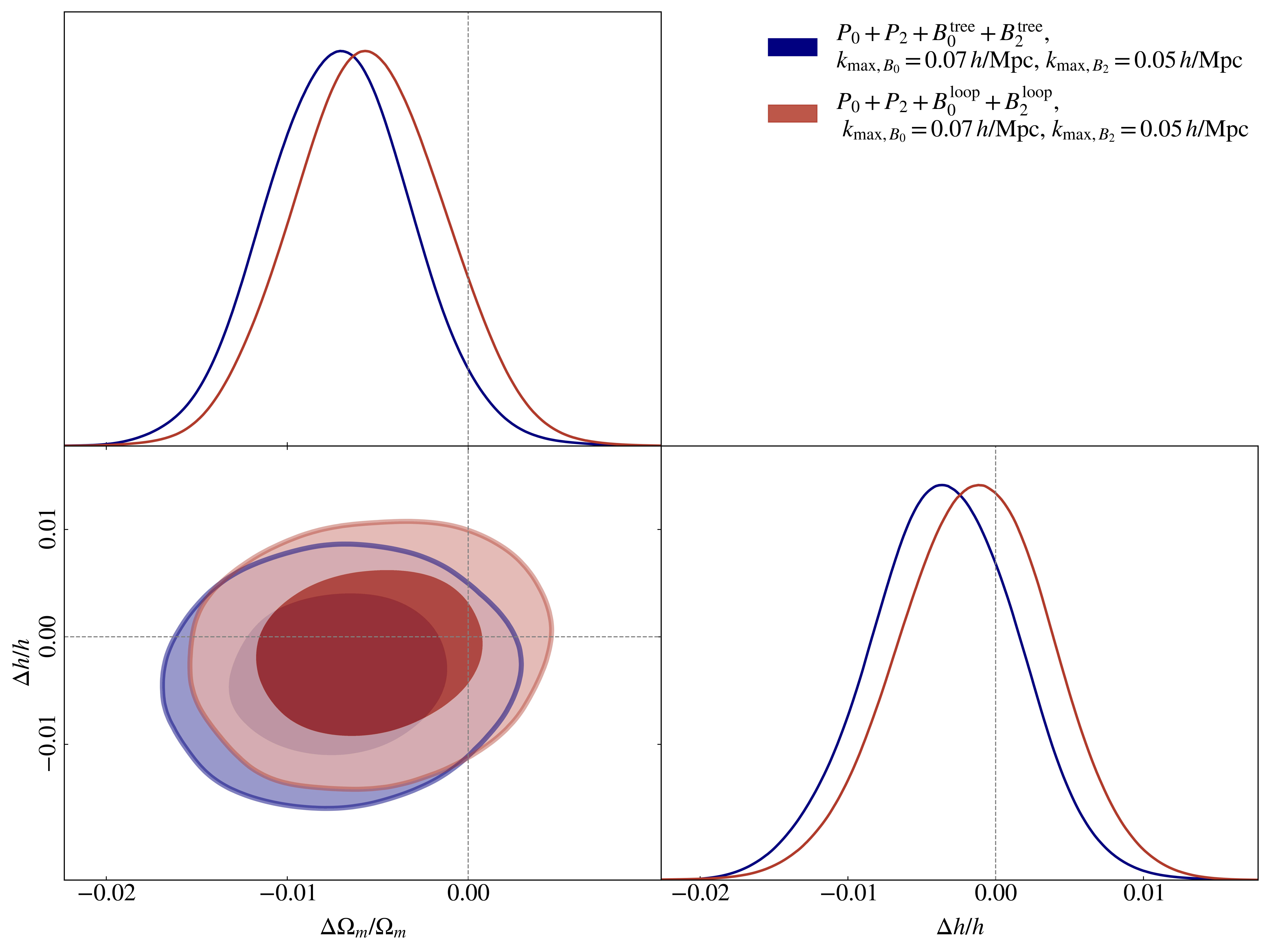}
    \caption{}  
    \label{fig:quad_loop_005}
  \end{subfigure}
 \caption{Impact of the 1-loop bispectrum (red) on cosmological constraints with respect to the tree-level contribution (blue). Analyses are performed at fixed cosmology ($\Lambda$CDM), with $n_s$ and $A_s$ fixed to the true values of the simulations.}
  \label{fig:bisp_1loop}
\end{figure}

In Fig.~\ref{fig:k_max_quad_cov_th}, we extend the analysis to include the bispectrum quadrupole, fix $k_{\text{max}, B_0}= 0.07 \,\ihMpc$, and examine how varying $k_{\text{max},B_2}$ impacts the results. We find that changing $k_{\text{max},B_2}$ has only a modest effect on the cosmological parameters and on $\varepsilon_f$, but it more noticeably alters the posterior of $\varepsilon_{d_{\gamma}}$. 
This pattern emerges because, as $k_{\text{max}}$ increases, 1-loop corrections and non-linear redshift-space distortions degrade the accuracy of our tree-level bispectrum description. This is illustrated in Fig.~\ref{fig:bisp_1loop}, where we compare $\Lambda$CDM analyses that include the 1-loop prediction for the bispectrum monopole and quadrupole with those based solely on the tree-level bispectrum multipoles, for $k_{\text{max}, B_2}=0.04 \ihMpc$ (left panel) and $k_{\text{max}, B_2}=0.05 \ihMpc$ (right panel). 

Conversely, lowering $k_{\text{max},B_2}$ below $0.04 \ihMpc$ significantly shrinks the number of usable triangular configurations, effectively removing the bispectrum quadrupole from the analysis. This can be verified by comparing the $\varepsilon_{d_{\gamma}}$ contours obtained for $k_{\text{max}, B_2}=0.03 \ihMpc$ in Fig.~\ref{fig:k_max_quad_cov_th} with those shown in Fig.~\ref{fig:k_max_mon_cov_th}. 

Taking all these results into account, we adopt $k_{\text{max},B_0} = 0.07 \,\ihMpc$ and $k_{\text{max},B_2} = 0.04 \,\ihMpc$ as our baseline choices for the PT Challenge analysis, as they provide a conservative yet unbiased setup for our tree-level bispectrum modeling.

\begin{comment}
\begin{table}[t]
    \centering
    \setlength{\tabcolsep}{8pt}
    \renewcommand{\arraystretch}{1.2} 
    \begin{minipage}{0.45\textwidth}
        \centering
        \begin{tabular}{|c|c|c|}
            \hline
            $k_{\text{max},B_0} \ [h/\text{Mpc}]$ & $\varepsilon_f$ & $\varepsilon_{d_\gamma}$ \\
            \hline
            0.06 & $-0.0018\pm0.0094$ & $0.61 \pm 0.37$  \\
            0.07 & $-0.0002\pm0.0097$ & $0.61 \pm 0.38$  \\
            0.08 & $0.0036\pm0.0097$ & $0.57 \pm 0.39$ \\
            \hline
        \end{tabular}
        \centering
    \end{minipage}
    \hfill
    \begin{minipage}{0.45\textwidth}
        \centering
        \begin{tabular}{|c|c|c|}
            \hline
            $k_{\text{max},B_2} \ [h/\text{Mpc}]$ & $\varepsilon_f$ & $\varepsilon_{d_\gamma}$ \\
            \hline
            0.03 & $-0.0005 \pm 0.0098$ & $0.55 \pm 0.35$ \\
            0.04 & $0.0071 \pm 0.0097$ & $-0.16 \pm 0.25$ \\
            0.05 & $0.0126 \pm 0.0088$ & $-0.89 \pm 0.17$ \\
            \hline
        \end{tabular}
        \centering
    \end{minipage}
    \caption{Constraints on the bootstrap parameters $\varepsilon_f$ and $\varepsilon_{d_\gamma}$ for different choices of $k_{\text{max},B_{\ell}}$ for the PT Challenge simulations. We fix $n_s$ and $A_s$ to the true values of the simulations. Reported values correspond to posterior means with $68\%$ confidence intervals.}
    \label{tab:kmax_eps_cov_th}
\end{table}
\end{comment}

%%%%%%%%%%%%%%%%%%%%%%%

\subsection{Degeneracies}
\label{sec:degeneracies}

\begin{figure}[H]
    \centering
    \includegraphics[width=0.7\linewidth]{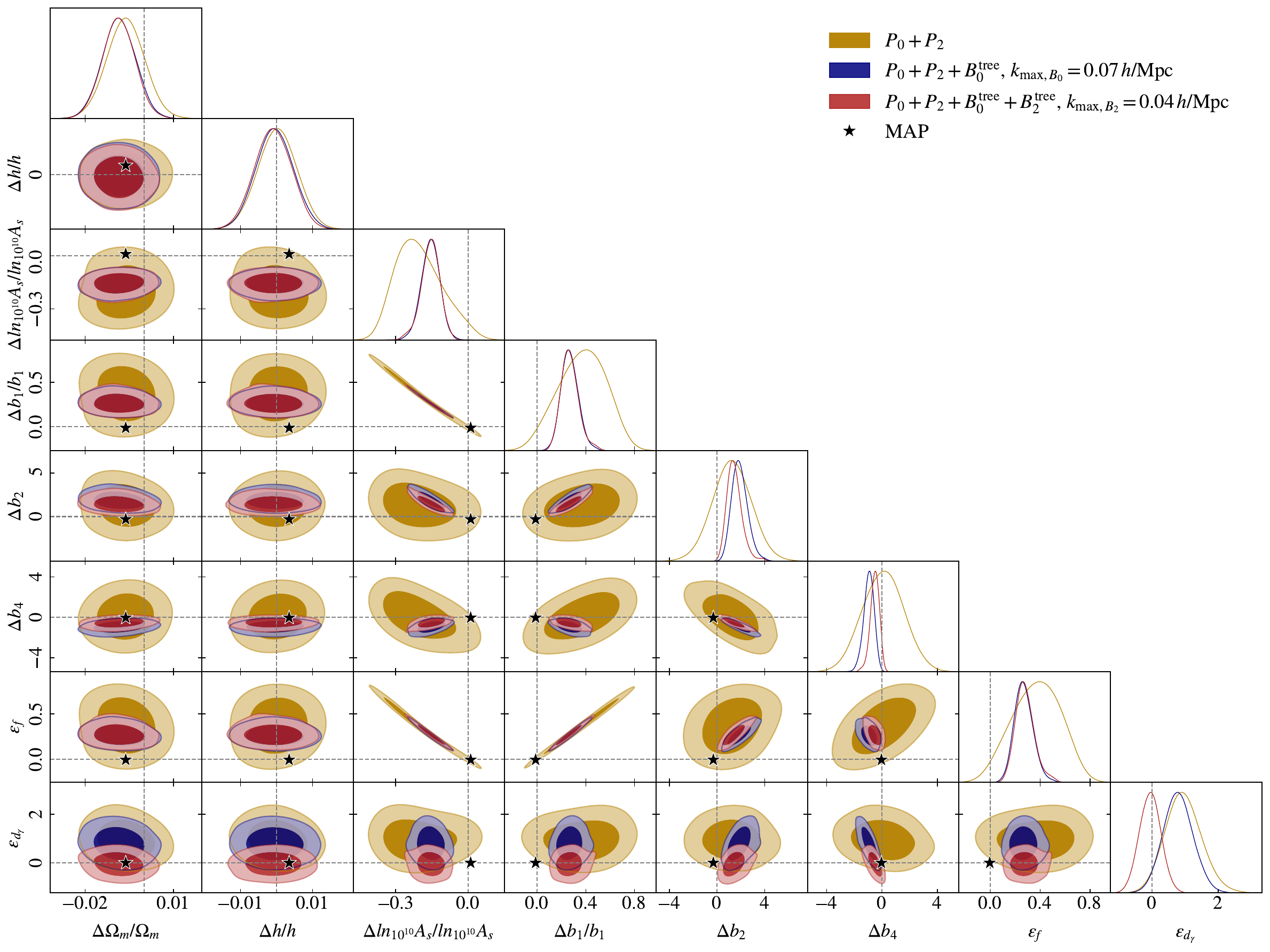}
    \caption{Posteriors for the PT Challenge simulations, with $n_s$ fixed to the true value and $A_s$ varied with a flat prior. We adopt $k_{\text{max},B_0}= 0.07 \ihMpc$ and $k_{\text{max},B_2}= 0.04 \ihMpc$. Dashed lines represent the fiducial values of the simulations for the cosmological and bootstrap parameters, and the inferred values for the bias ones (see Appendix \ref{app:full_posteriors} for details). The shifts of the posteriors with respect to dashed lines indicate biased results when $\varepsilon_f$, $A_s$ and $b_1$ are allowed to vary simultaneously. Stars indicate the MAP values obtained from minimization.}
    \label{fig:degeneracies} 
\end{figure}

In this section, we examine the parameter degeneracies that impact our analysis. 
At linear order, it is well known that the amplitude of primordial scalar fluctuations, $A_s$, is completely degenerate with the linear galaxy bias, $b_1$. This degeneracy is broken in redshift space, in absence of strong observational effects~\cite{Desjacques:2018pfv}, assuming a perfect knowledge of the growth function $f$. Leaving $f$ free to vary in the model, as we do in this work in the spirit of a model-agnostic analysis, will restore the broken degeneracy, and cause possible projection effects in the parameter posterior. As shown in Fig.~\ref{fig:degeneracies}, the degeneracies among $b_1$, $A_s$ and $f$ are partially broken once nonlinear effects are taken into account, both through the 1-loop corrections to the power spectrum and through the inclusion of the bispectrum, as well as by exploiting information from redshift-space anisotropies, including higher-order multipoles.

Nonetheless, when all three parameters are allowed to vary simultaneously, projections along these degenerate directions in parameter space yield posterior distributions that are shifted relative to the fiducial values. This behavior is a generic feature of models beyond $\Lambda$CDM, as previously emphasized in~\cite{Piga:2022mge, Carrilho:2022mon, Simon:2022lde, Taule:2024bot}.

\begin{figure}[H]
    \centering
    \includegraphics[width=0.6\linewidth]{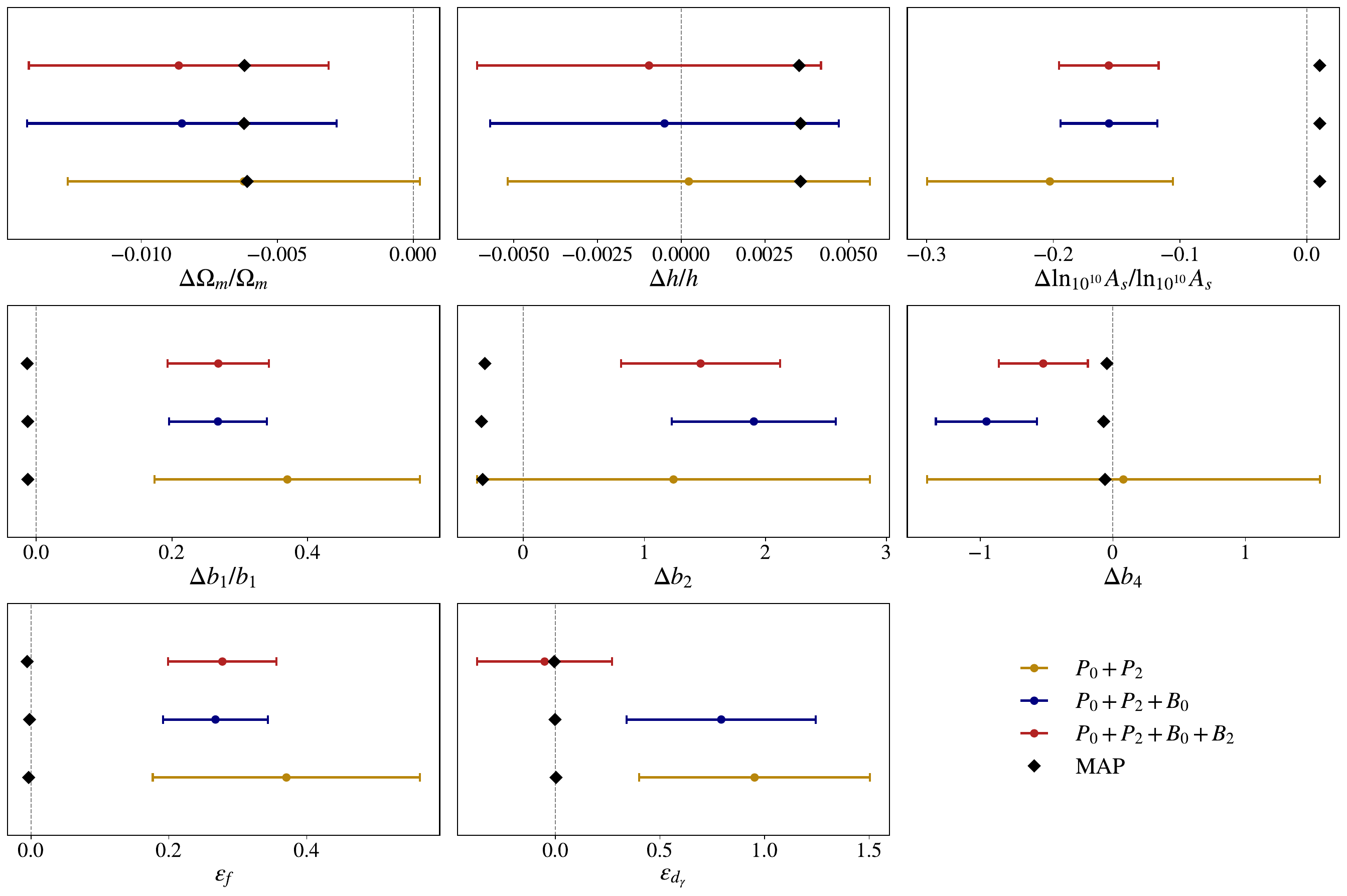}
    \caption{Posterior means and $1\sigma$ intervals (colored bars) and MAP values, denoted by black diamond markers, for the PT Challenge simulations, with $A_s$ free. Projection effects are reflected in the discrepancies between the MAP values and the maxima of the marginalized posterior distributions.}
    \label{fig:projection_effects} 
\end{figure}

To quantify these projection effects, following~\cite{Simon:2022lde}, we compare the MAP values obtained from minimization with the 68\%-credible intervals in Fig.~\ref{fig:degeneracies}. As shown in Fig.~\ref{fig:projection_effects}, the inclusion of bispectrum multipoles, compared to the power spectrum only case, provides only a marginal reduction of the projection effects on $A_s$, $b_1$ and $\varepsilon_f$, which remain large even when including the bispectrum quadrupole $B_2$ ($\sim 4.23\sigma$, $3.78\sigma$ and $3.61\sigma$, respectively). This confirms that the $A_s$--$b_1$--$\varepsilon_f$ degeneracy cannot be resolved by the bispectrum multipoles alone. 

Since the main objective of our work is to constrain signatures of new physics, rather than to perform a full $\Lambda$CDM analysis based solely on galaxy clustering, we fix $A_s$ to the true value used in the PT Challenge simulations. This prescription effectively mimics the inclusion of external information from complementary probes, such as the CMB, as also shown in Appendix~\ref{app:As_comparison}. Accordingly, in what follows we present results with $A_s$ fixed to the true value of the PT Challenge, a choice that guarantees convergence and enables a robust inference of cosmological, bias, and model-independent parameters.

In contrast to $\varepsilon_f$, the bootstrap parameter $\varepsilon_{d_{\gamma}}$ does not show significant degeneracies with the cosmological parameters, but it is degenerate with the quadratic bias coefficients $b_2$ and $b_4$. This can be understood by looking at Eq.~\eqref{eq:b24K}, which shows the relation between $b_{2,4}$ and $b_{K^2}$, and noticing the relation between $\gamma$ and $K^2$ is $\gamma(\bq_1, \bq_2) = 2/3 - K^2(\bq_1, \bq_2)$. This means that a better knowledge of higher order bias parameters (second order in this specific case) will substantially increase the sensitivity to $\varepsilon_{d_\gamma}$. This is what we observe in our analysis: including the bispectrum quadrupole improves the constraints on this beyond-$\Lambda$CDM parameter. In particular, the projection effects on $\varepsilon_{d_\gamma}$ decrease from $1.76\sigma$ ($P_0+P_2+B_0$) to $0.14\sigma$ ($P_0+P_2+B_0+B_2$), and a similar reduction is observed for the bias parameter $b_4$ (from $2.31\sigma$ to $1.43\sigma$), consistently with the degeneracy between $\varepsilon_{d_\gamma}$ and the quadratic bias parameters discussed above. Consequently, the bootstrap parameter $\varepsilon_{d_{\gamma}}$ can be tightly constrained even without fixing the amplitude of primordial scalar fluctuations, $A_s$, making it a robust probe of new physics beyond $\Lambda$CDM. In contrast, the projection effects on $b_2$ remain large even with the inclusion of the bispectrum quadrupole, since $b_2$ is strongly correlated with $b_1$ through Eq.~\eqref{eq:b24K}. 

%\begin{table}[t]
%    \centering
%    \resizebox{\textwidth}{!}{%
%    \setlength{\tabcolsep}{12pt}
%    \renewcommand{\arraystretch}{1.2}
%    \begin{tabular}{|c|c|c|c|c|c|c|}
%        \hline
%        \textbf{Dataset} & $\Delta\Omega_m/\Omega_m$ & $\Delta h/h$ & $\Delta \ln{(10^{10}A_s)}/\ln{(10^{10}A_s)}$ & $\Delta b_1/b_1$ & $\varepsilon_f$ & $\varepsilon_{d_\gamma}$ \\
%        \hline
%        $P_l$ & $ -0.0062 \pm 0.0065 $  & $ 0.0002 \pm 0.0054 $ & $ -0.205^{+0.077}_{-0.11} $ & $ 0.37^{+0.21}_{-0.18}$ & $ 0.37^{+0.21}_{-0.19} $ & $ 0.95^{+0.51}_{-0.59} $   \\
%        $P_l+B_0$ & $ -0.0085 \pm 0.0057 $ & $ -0.0004 \pm 0.0053$ & $ -0.157^{+0.040}_{-0.036}$ & $ 0.270^{+0.064}_{-0.078}$ & $ 0.27^{+0.066}_{-0.083}$ & $ 0.79 \pm 0.45 $ \\
%        $P_l+B_0+B_2$ & $-0.0087 \pm 0.0056 $ & $-0.0011 \pm 0.0052 $  & $ -0.155 \pm 0.040 $ & $0.266^{+0.067}_{-0.076} $ & $0.275^{+0.070}_{-0.081} $ & $-0.05\pm0.32 $  \\
%        \hline
%    \end{tabular}
%    } 
%    \caption{Constraints for the PT Challenge simulations, with $n_s$ fixed and $A_s$ varied with a flat prior. Reported values correspond to posterior means with $68\%$ confidence intervals.}
%    \label{tab:res_As_free_cov_th}
%\end{table}

In the BOSS analysis, the degeneracies are less pronounced because of the smaller survey volume and the correspondingly broader constraints. For this reason, in this case we show results both allowing $A_s$ to vary and keeping it fixed; see Sect.~\ref{sec:BOSS_res} and Fig.~\ref{fig:BOSS_res}.

%%%%%%%%%%%%%%%%%%%%%%%
\section{Results}
\label{sec:results}
%\subsection{BOSS analysis: $A_s$ free}
\subsection{BOSS analysis}
\label{sec:BOSS_res}

\subsubsection{$A_s$ free}

Figure~\ref{fig:res_BOSS_As_free} shows the marginalized posterior distributions of the cosmological and bootstrap parameters for the BOSS LOWZ and CMASS samples, with the corresponding constraints summarized in the upper block of Table~\ref{tab:res_BOSS_As_combined}.

The power spectrum multipoles on their own do not offer enough constraining power to jointly determine the cosmological parameters and the new model-independent parameters $\varepsilon_f$ and $\varepsilon_{d_{\gamma}}$. As highlighted in the previous section, the strong degeneracy between $\varepsilon_f$ and $A_s$ induces projection effects and biased constraints, preventing us from recovering values consistent with those derived from \textit{Planck} data \cite{Planck2018}. 

For this reason, incorporating the bispectrum is essential: it greatly enhances both the precision and the overall constraining ability of the analysis. This agrees with earlier works that combine power spectrum and bispectrum (P+B) \cite{Rizzo:2022lmh,Tsedrik:2022cri, Carrilho:2022mon, DAmico:2022osl, Ivanov:2021kcd,Ivanov:2023qzb}, which have shown that the bispectrum provides extra constraining power, helping to break parameter degeneracies and reduce projection effects.

Focusing on the bootstrap parameters, we observe that incorporating the bispectrum leads to posterior distributions of $\varepsilon_f$ that are consistent with its $\Lambda\text{CDM}$ prediction and decreases its uncertainty from $0.24$ to about $0.135$ in LOWZ and from $0.16$ to $0.097$ in CMASS, corresponding to reductions of $44\%$ and $39\%$, respectively. The parameter $\varepsilon_{d_{\gamma}}$ is already in agreement with $\Lambda\text{CDM}$ when only the power spectrum is considered; nonetheless, adding the bispectrum monopole further sharpens its constraints, cutting the uncertainty from $1.3$ in both redshift bins down to $0.69$ (LOWZ) and $0.68$ (CMASS), which represents an improvement of roughly $47\%$.

%\begin{figure}[t]
%    \centering
%    \includegraphics[width=0.65\linewidth]{img/BOSS_As_free.jpg}
%    \caption{Marginalized posterior distributions for the cosmological and model-independent parameters from the joint BOSS LOWZ and CMASS analysis, with $n_s$ fixed to the \textit{Planck} best-fit value. Dashed lines indicate the fiducial \textit{Planck} values \cite{Planck2018}.}
%    \label{fig:res_BOSS_As_free}
%\end{figure}

%\begin{table}[H]
%\centering
%\setlength{\tabcolsep}{7pt}
%\renewcommand{\arraystretch}{1.2}
%\begin{tabular}{|c|c|c|c|cc|cc|}
%\hline
%\textbf{Dataset} & $\Omega_m$ & $h$ & %$\ln(10^{10}A_s)$ &
%\multicolumn{2}{c|}{$\varepsilon_f$} & 
%\multicolumn{2}{c|}{$\varepsilon_{d_{\gamma}}$} \\
%\hline
% & & & & LOWZ & CMASS & LOWZ & CMASS \\
%\hline
%$P_l$ & $0.3071 \pm 0.0099$ & $0.690 \pm 0.011$ & %$1.87^{+0.14}_{-0.19}$ & $0.90 \pm 0.24$ & $0.56 %\pm 0.16$ & $-0.1 \pm 1.3$ & $1.0 \pm 1.3$ \\
%$P_l+B_0$ & $0.307 \pm 0.011$ & %$0.696^{+0.011}_{-0.013}$ & $3.00^{+0.19}_{-0.17}$ %& $0.19^{+0.12}_{-0.15}$ & $-0.015 \pm 0.097$ & %$-0.38 \pm 0.69$ & $0.27 \pm 0.68$ \\
%hline
%\end{tabular}

%\caption{Constraints on cosmological and model-independent parameters from the BOSS LOWZ and CMASS datasets with $n_s$ fixed. Cosmological parameters are shared between the two redshift bins, while the model-independent parameters $\varepsilon_f$ and $\varepsilon_{d_{\gamma}}$, being time dependent, are independently fitted for each sample. Reported values correspond to posterior means with 68\% confidence intervals.}
%\label{tab:res_BOSS_As_free}
%\end{table}

\subsubsection{$A_s$ fixed}
We re-ran the BOSS analysis with $A_s$ fixed to the \textit{Planck} best-fit value \cite{Planck2018}. This choice enables a consistent comparison between the BOSS and PT Challenge results and demonstrates how a joint analysis of data from forthcoming galaxy surveys together with external datasets can further tighten current constraints on model-independent parameters. %It is also essential to reach convergence of the MCMC chains.

\begin{figure}[H]
  \centering
  \begin{subfigure}{0.75\textwidth}
    \includegraphics[width=\linewidth]{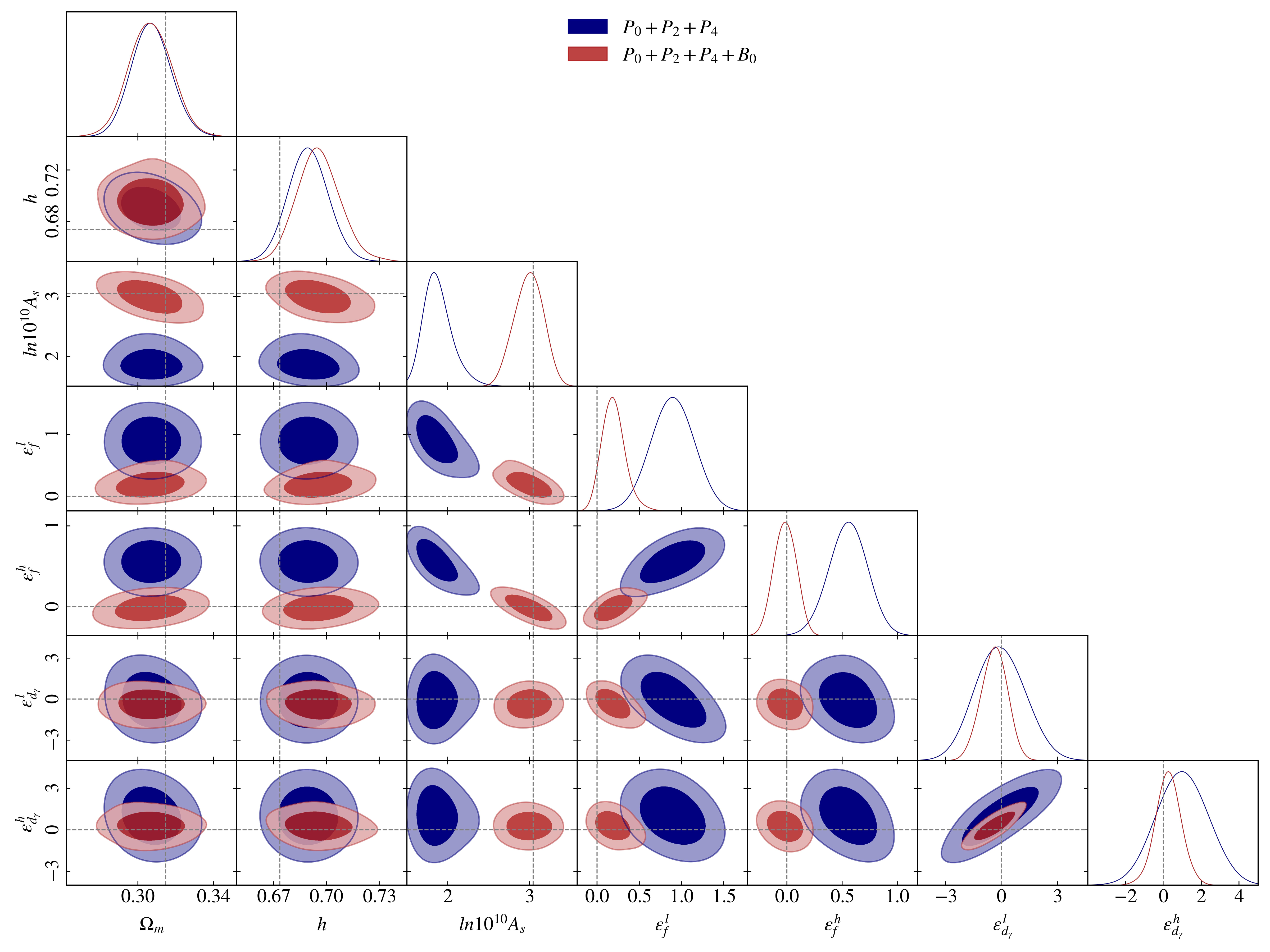}
    \caption{}
    \label{fig:res_BOSS_As_free}   
  \end{subfigure}
  \hfill
  \begin{subfigure}{0.75\textwidth}
    \includegraphics[width=\linewidth]{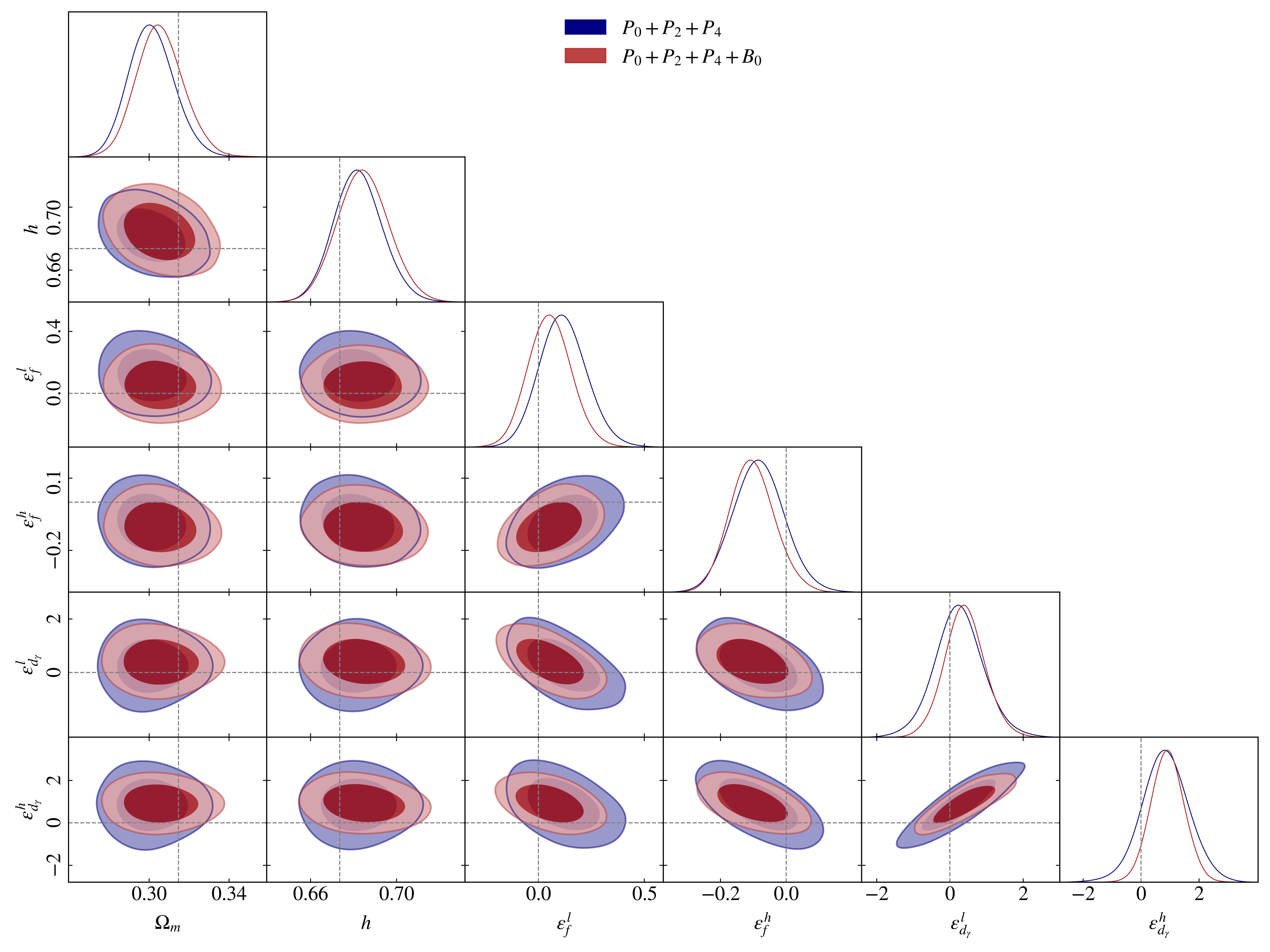}
    \caption{}
    \label{fig:res_BOSS_As_fixed}
  \end{subfigure}
  \caption{Marginalized posterior distributions for the BOSS LOWZ and CMASS datasets with $n_s$ fixed to \textit{Planck} best-fit and a flat prior on $A_s$ (top panel) and $A_s$ fixed to \textit{Planck} central value (bottom panel). Dashed lines represent the best-fit values  from \textit{Planck} \cite{Planck2018}.}
  \label{fig:BOSS_res}
\end{figure}

As illustrated in Fig.~\ref{fig:res_BOSS_As_fixed}, when $A_s$ is held fixed, the bispectrum does not add substantial constraining power beyond that of the power spectrum alone. It does, however, lead to a modest tightening of the bootstrap parameter bounds. In particular, the uncertainties on $\varepsilon_f$ are further reduced by about $25\%$ for LOWZ and $20\%$ for CMASS, while those on $\varepsilon_{d_{\gamma}}$ decrease by roughly $20\%$. The corresponding parameter constraints are reported in the lower block of Table~\ref{tab:res_BOSS_As_combined}.

\begin{table}[H]
\centering
\setlength{\tabcolsep}{5pt}
\renewcommand{\arraystretch}{1.2}
\begin{tabular}{|c|c|c|c|c|cc|cc|}
\hline
\multirow{2}{*}{} 
& \multirow{2}{*}{\centering \textbf{Dataset}} 
& \multirow{2}{*}{\centering $\Omega_m$} 
& \multirow{2}{*}{\centering $h$} 
& \multirow{2}{*}{\centering $\ln(10^{10}A_s)$} 
& \multicolumn{2}{c|}{$\varepsilon_f$} 
& \multicolumn{2}{c|}{$\varepsilon_{d_{\gamma}}$} \\
\cline{6-9}
& & & & & LOWZ & CMASS & LOWZ & CMASS \\
\hline
\multirow{2}{*}{\textbf{$A_s$ free}} 
& $P_l$ 
& $0.3071 \pm 0.0099$ 
& $0.690 \pm 0.011$ 
& $1.87^{+0.14}_{-0.19}$ 
& $0.90 \pm 0.24$ 
& $0.56 \pm 0.16$
& $-0.1 \pm 1.3$ 
& $1.0 \pm 1.3$ \\
& $P_l+B_0$ 
& $0.307 \pm 0.011$ 
& $0.696^{+0.011}_{-0.013}$ 
& $3.00^{+0.19}_{-0.17}$  
& $0.19^{+0.12}_{-0.15}$ 
& $-0.015 \pm 0.097$ 
& $-0.38 \pm 0.69$ 
& $0.27 \pm 0.68$ \\
\hline
\hline
\multirow{2}{*}{\textbf{$A_s$ fixed}} 
& $P_l$ 
& $0.301^{+0.011}_{-0.012}$ 
& $0.682^{+0.011}_{-0.012}$ 
& -- 
& $0.12^{+0.10}_{-0.12}$  
& $-0.084 \pm 0.078$ 
& $0.25 \pm 0.67$ 
& $0.84 \pm 0.82$ \\
& $P_l+B_0$ 
& $0.306^{+0.011}_{-0.012}$ 
& $0.685 \pm 0.012$ 
& -- 
& $0.05 \pm 0.10$ 
& $-0.103^{+0.064}_{-0.072}$ 
& $0.40 \pm 0.54$ 
& $0.91 \pm 0.57$ \\
\hline
\end{tabular}
\caption{Constraints on cosmological and model-independent parameters derived from the BOSS LOWZ and CMASS samples. Top panel: $A_s$ allowed to vary. Bottom panel: $A_s$ held fixed. The cosmological parameters are common to both redshift bins, whereas the model-independent parameters $\varepsilon_f$ and $\varepsilon_{d_{\gamma}}$, which depend on time, are fit separately for each sample. The quoted values are posterior means with 68\% confidence intervals.}
\label{tab:res_BOSS_As_combined}
\end{table}

\subsection{PT Challenge simulations}
\label{sec:PT_Challenge_res_cov_th}

The BOSS analysis highlights the advantages of combining the power spectrum with the bispectrum, but the limited survey volume restricts how precisely we can determine the model-independent parameters $\varepsilon_f$ and $\varepsilon_{d_\gamma}$. To assess the potential gains achievable with larger datasets, we therefore turn to the PT Challenge simulations, whose significantly greater volume enables much tighter constraints.

\begin{figure}[H]
    \centering
    \includegraphics[width=0.75\linewidth]{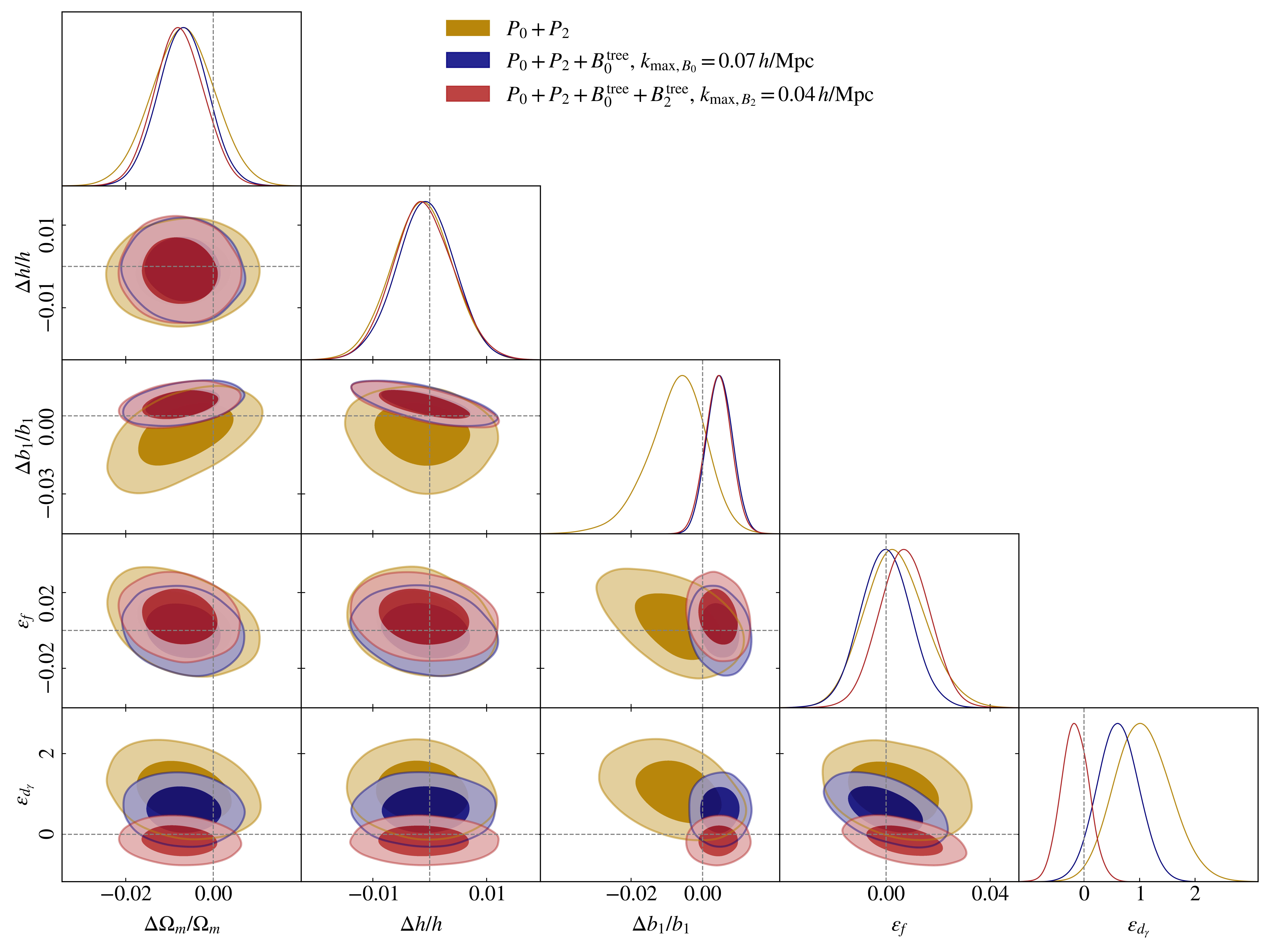}
    \caption{Posteriors for the PT Challenge, with $n_s$ and $A_s$ fixed to the truth of the simulations. We adopt $k_{\text{max},B_0}= 0.07 \ihMpc$ and $k_{\text{max},B_2}= 0.04 \ihMpc$.}
    \label{fig:As_fixed_results_th_cov}
\end{figure}

Figure \ref{fig:As_fixed_results_th_cov} displays the marginalized posterior distributions for the cosmological parameters, and the corresponding numerical constraints are summarized in Table \ref{tab:res_As_fixed_th_cv}.  
When the bispectrum multipoles are included, the constraints on $\varepsilon_f$ are only modestly improved relative to those obtained from the power spectrum alone, tightening from about $0.012$ to $0.0097$. In contrast, adding the bispectrum quadrupole is crucial for obtaining unbiased constraints on $\varepsilon_{d_{\gamma}}$, yielding uncertainties at the $\sim 0.25$ level.

Furthermore, in comparison with the $A_s$-free analysis shown in Fig.~\ref{fig:degeneracies}, fixing $A_s$ allows us to accurately recover the fiducial PT Challenge values of both the linear bias $b_1$, the quadratic biases, $b_2$ and $b_4$\footnote{The value of the simulation biases parameter $b_1$, $b_2$ and $b_4$ have been estimated via the peak-background split approach, as described in Appendix \ref{app:full_posteriors}.} and the model-independent parameter $\varepsilon_f$.

Relative to the BOSS results, the increase in constraining power provided by the PT Challenge data is dramatic. Since these simulations cover a volume roughly 100 times larger, they reduce the uncertainty on $\varepsilon_f$ by a factor of 7, improving the precision from about $0.068$ in the high-redshift BOSS sample to $0.01$ in the PT Challenge. Additionally, the parameter $\varepsilon_{d_{\gamma}}$, which BOSS could only weakly constrain, now has an uncertainty reduced by a factor of about 2.5, down to the $\sim 0.25$ level.

\begin{table}[H]
    \centering
    \setlength{\tabcolsep}{12pt}
    \renewcommand{\arraystretch}{1.2} 
    \begin{tabular}{|c|c|c|c|c|}
        \hline
        \textbf{Dataset} &$\Delta\Omega_m/\Omega_m$ & $\Delta h/h$  & $\varepsilon_f$ & $\varepsilon_{d_\gamma}$ \\
        \hline
        $P_l$ & $ - 0.0069 \pm 0.0071 $ & $ -0.0014 \pm 0.0054 $ & $ 0.003 \pm 0.012$ & $ 1.06^{+0.47}_{-0.53} $ \\
        $P_l+B_0$ & $ -0.0068 \pm 0.0056 $ & $ -0.0008 \pm 0.0052$ & $ -0.0002 \pm 0.0097 $ & $ 0.61 \pm 0.38 $ \\
        $P_l+B_0+B_2$ & $ -0.0077 \pm 0.0057$ & $-0.0011 \pm 0.0053$ & $0.0071 \pm 0.0097 $& $-0.16 \pm 0.25 $  \\
        \hline
    \end{tabular}
    \caption{Constraints for the PT Challenge simulations, with $n_s$ and $A_s$ fixed. Reported values correspond to posterior means with $68\%$ confidence intervals.}
    \label{tab:res_As_fixed_th_cv}
\end{table}

%%%%%%%%%%%%%%%%%%%%%%%%%%%%%%%%%%%%%%%%%%%%%
\section{Conclusion and outlook}
\label{sec:conclusions}

In this work, we have employed a model-independent framework based on the LSS bootstrap to probe departures from the standard $\Lambda$CDM picture over a broad class of cosmological models. 
The primary objective of our study has been to determine how tightly current and forthcoming galaxy clustering measurements can constrain both the linear and nonlinear perturbation sectors, characterized by the bootstrap parameters $\varepsilon_f$ and $\varepsilon_{d_{\gamma}}$.

By jointly analyzing $\varepsilon_f$, which encodes deviations in the linear growth, and $\varepsilon_{d_{\gamma}}$, which describes modifications in the nonlinear evolution, we can alleviate degeneracies between different extensions of the $\Lambda$CDM framework.
Our study makes use of the \texttt{PyBird} code, which we have adapted to consistently incorporate the LSS Bootstrap parametrization and the exact time evolution in the $\Lambda$CDM limit. We have implemented this pipeline on both observational data from the BOSS survey and mock data from the large-volume PT Challenge simulations.
Using a state-of-the-art P+B analysis, with the power spectrum modeled up to one-loop and the bispectrum multipoles evaluated at tree-level, we obtain constraints on the linear $\varepsilon_f$ of $\sim 0.07 \ (0.10)$ level and on the bootstrap parameter $\varepsilon_{d_{\gamma}}$ at the $\sim 0.57 \ (0.54)$ level for the CMASS (LOWZ) samples. For the PT Challenge simulations, we find significantly tighter constraints, reaching the $\sim 0.01$ level on $\varepsilon_f$ and $\sim 0.25$ on $\varepsilon_{d_{\gamma}}$.
Our findings underscore the crucial impact of bispectrum multipoles on constraining the nonlinear parameter $\varepsilon_{d_{\gamma}}$: in particular, for the PT Challenge dataset, the redshift-space information encoded in the bispectrum quadrupole adds information that tightens the constraints on $\varepsilon_{d_\gamma}$ and also reduces its projection effects.

The marked gains exhibited by the PT Challenge simulations highlight the pivotal role of survey volume in tightening constraints on the bootstrap parameters. However, assuming good control of systematic errors, the ultimate improvement is limited by the reliability of the theoretical framework, most notably the modeling of redshift-space distortions, which sets the minimum scales that can be robustly exploited in the analysis.  
Accordingly, a natural next step is to include the bispectrum at one-loop order, thereby allowing a self-consistent extension of the analysis to larger values of $k_{\text{max}}$.

The key question at this stage is whether the current ‘state‑of‑the‑art’ constraints obtained in this work can be sharpened by incorporating information beyond the power spectrum and the (potentially one‑loop) bispectrum, either through alternative summary statistics or through fully field‑level analyses (see, for example, \cite{Beyond-2pt:2024mqz, Peron:2024xaw, Marinucci:2024bdq, Nguyen:2024yth, Spezzati:2025zsb, Akitsu:2025boy}).

An initial step toward realizing field‑level inference within the bootstrap framework was presented in \cite{Peron:2025lgh}, which examined the matter distribution in real space for fixed initial conditions. To push this approach further and apply it to actual observations, sophisticated sampling techniques such as Hamiltonian Monte Carlo will be needed to marginalize over both the amplitudes and phases of the linear modes.

We leave these interesting developments for future work.

\section*{Acknowledgments}
\noindent
G.~Bizelli and M.~Pietroni thank the Yukawa Institute for Theoretical Physics (Kyoto), where this work was completed, for hospitality.
We thank Takahiro Nishimichi for providing us with the data from the `PT Challenge' simulations and assistance on how to use them, and Matteo Peron for collaborating to the initial stages of this project. We thank Kevin Pardede for useful discussions regarding the calculations of the theoretical covariance.

Numerical computations were performed on the High Performance Computing facility of the University of Parma, Italy, whose support team we thank.

\appendix

\section{Equations of motion and Perturbation Theory Kernels} 
\label{app:kernel}

The Fourier space equations of motion for the dark matter overdensity $\delta$ and the velocity divergence $\theta$ are given by

\begin{align}
    &\partial_\eta\delta_{\bk} -f\theta_{\bk} = f \, \mathcal{I}_{\bk, \bq_1, \bq_2}\alpha(\bq_1, \bq_2) \theta_{\bq_1}\delta_{\bq_2} \ , \nonumber \\
    &\partial_\eta\theta_{\bk} - f \theta_{\bk} + \frac{3}{2}\frac{\Omega_m}{f}\mu_{\Phi} \theta_{\bk} + \frac{1}{f}\frac{k^2}{\mathcal{H}^2}\Phi_{\bk} = f ,\ \mathcal{I}_{\bk, \bq_1, \bq_2}\beta(\bq_1, \bq_2) \theta_{\bq_1} \theta_{\bq_2} \ ,
    \label{eq_motion}
    \end{align}

where the time dependence has been omitted to avoid clutter, and we have defined 

\begin{equation}
    \mathcal{I}_{\bk, \bq_1, \dots, \bq_n} \equiv \int \frac{d^3\bq_1}{(2\pi)^3} \dots \frac{d^3\bq_n}{(2\pi)^3} \delta_D(\bk - \bq_{1 \dots n}) \, ,
\end{equation}

with $\bq_{1 \dots n} = \sum_{i=1}^n \bq_i$. The functions $\alpha$ and $\beta$ are the standard dark matter interaction vertices, that describe the non-linear coupling between different modes in the evolution of the density and velocity fields, defined as

\begin{align}
    \alpha(\bq_1, \bq_2) \equiv \frac{\bq_1 \cdot (\bq_1+\bq_2) }{q_1^2} \ , \qquad
    \beta(\bq_1, \bq_2) \equiv \frac{|\bq_1+\bq_2|^2 \bq_1 \cdot \bq_2}{2 q_1^2 q_2^2}.
\end{align}

The time-dependent function $\mu_{\Phi}(\eta)$ encodes possible linear modifications to the gravitational interaction. General relativity is recovered for $\mu_{\Phi} = 1$.

The system is closed by the Poisson equation for the gravitational potential $\Phi$, given by

\begin{equation}
    - \frac{k^2}{\mathcal{H}^2}\Phi_{\bk} = \frac{3}{2}\mu_{\Phi} \Omega_m \delta_{\bk}.
\end{equation}

Possible non-linear modifications would enter through higher-order terms.

We proceed by summarizing the explicit expression for the perturbation theory kernels up to third order, following \cite{Piga:2022mge}\footnote{With respect to this reference we have defined \begin{equation*}
    d_{\gamma} = d_{\gamma}^{(2)} \ , \qquad d_{\gamma \gamma} = \frac{1}{4} d_{\gamma a}^{(3)} + d_{\gamma b}^{(3)} \ ,  \qquad d_{\gamma \alpha} = \frac{1}{4} d_{\gamma a}^{(3)} - \frac{1}{2} d_{\gamma b}^{(3)}.
\end{equation*}}. The galaxy kernels are 

\begin{align}
    K_1(\bq_1, a) & = b_1 \ , \\
    K_2(\bq_1,\bq_2, a) & = (-b_1+b_2+b_4) + b_1 \beta(\bq_1,\bq_2) + \bigg(b_1 - \frac{2}{7} b_2 \bigg) \gamma(\bq_1, \bq_2) \ ,  \\
    K_3(\bq_1,\bq_2,\bq_3, a)\big|_{\text{sub}} & = \frac{b_1}{3}O_{\beta \beta}(\bq_1,\bq_2,\bq_3) + \frac{1}{3} \bigg(\frac{g(a)b_1}{2} + \frac{b_3}{21} \bigg) O_{\gamma \beta}(\bq_1,\bq_2,\bq_3) \nonumber \\
    & + \frac{1}{3} \bigg(\frac{g(a)b_1}{2} - \frac{b_3}{21} \bigg) \bigg( O_{\gamma \gamma} (\bq_1,\bq_2,\bq_3) + \frac{1}{2} O_{\gamma \alpha_a}(\bq_1,\bq_2,\bq_3) \bigg) + \text{cyclic} \ , 
\end{align}

where the subscript ``sub'' denotes the subtracted (finite) part of the third-order kernel in the limit $q/k \to \infty$. This corresponds to the contribution that enters the 1-loop power spectrum \eqref{ps_1loop} through the particular combination of momenta $K_3(\bk, \bq, -\bq)$.

The velocity field kernels are given by

\begin{align}
    G_1(\q_1, a) & = 1 \ ,  \\
    G_2(\bq_1, \bq_2, a) & = \beta(\bq_1, \bq_2) + \frac{d_{\gamma}^{(2)}}{2}\gamma(\bq_1,\bq_2) \ ,  \\
    G_3(\bq_1, \bq_2, \bq_3, a) & = \frac{1}{6} \bigg[ 2 O_{\beta \beta}(\bq_1, \bq_2, \bq_3) + \big[ 2 d_{\gamma}^{(2)} + d_{\gamma a}^{(3)} - 2 (d_{\gamma a}^{(3)}+ g(a) )\big] O_{\beta \gamma}(\bq_1, \bq_2, \bq_3) \nonumber \\
    & + \bigg(\frac{1}{4}d_{\gamma a}^{(3)} - \frac{1}{2}d_{\gamma b}^{(3)} \bigg) O_{\gamma \alpha_a}(\bq_1, \bq_2, \bq_3) + \bigg(d_{\gamma b}^{(3)} - \frac{1}{2}d_{\gamma a}^{(3)} + 2g(a) \bigg) O_{\gamma \beta}(\bq_1, \bq_2, \bq_3) \nonumber \\
    & + \bigg(\frac{1}{2}d_{\gamma a}^{(3)} + d_{\gamma b}^{(3)} \bigg) O_{\gamma \gamma}(\bq_1, \bq_2, \bq_3) + \text{cyclic} \bigg]\ ,
\end{align}

where we have defined

\begin{align}
    O_{\beta \beta}(\bq_1, \bq_2, \bq_3 ) & \equiv \beta(\bq_1, \bq_2) \beta(\bq_{12}, \bq_3 ) \ , \nonumber \\
    O_{\beta \gamma}(\bq_1, \bq_2, \bq_3 ) & \equiv \beta(\bq_1, \bq_2) \gamma(\bq_{12}, \bq_3 ) \ , \nonumber \\
    O_{\gamma \beta }(\bq_1, \bq_2, \bq_3 ) & \equiv \gamma(\bq_1, \bq_2) \beta(\bq_{12}, \bq_3 ) \ , \nonumber \\
    O_{\gamma \gamma }(\bq_1, \bq_2, \bq_3 ) & \equiv \gamma(\bq_1, \bq_2) \gamma(\bq_{12}, \bq_3 ) \ , \nonumber \\
    O_{\gamma \alpha_a }(\bq_1, \bq_2, \bq_3 ) & \equiv \gamma(\bq_1, \bq_2) \alpha_a(\bq_{12}, \bq_3 ) \ ,
\end{align}

and

\begin{align}
    & \alpha_a(\bq_1, \bq_2) \equiv \alpha(\bq_1, \bq_2) - \alpha(\bq_2, \bq_1) \ , \nonumber \\
    & g(a) \equiv \int_0^a d\ln{\tilde{a}}f(\tilde{a}) \bigg[\frac{D(\tilde{a})}{D(a)} \bigg]^2 d_{\gamma}^{(2)}(\tilde{a}).
\end{align}

In the EdS approximation, the time-dependent coefficients in the kernels above take the constant values

\begin{align}
    d_{\gamma,\text{EdS}}^{(2)} = \frac{6}{7}\ , \qquad d_{\gamma a,\text{EdS}}^{(3)} = \frac{2}{7}\ ,  \qquad d_{\gamma b,\text{EdS}}^{(3)} = \frac{5}{21}.
\end{align}

The redshift space galaxy kernels are built as usual from the ones for density and velocity

\begin{align}
    Z_1(\bk, a) & =  b_1+f\mu_k^2 \ , \\
    Z_2(\bq_1,\bq_2, a)  & =  K_2(\bq_1,\bq_2)+ f \mu^2_k G_2(\bq_1,\bq_2) + b_1 f \mu_k k\Bigg(\frac{\mu_{q_1}}{q_1}+\frac{\mu_{q_2}}{q_2}\Bigg) + f^2\mu_k^2 k^2 \frac{\mu_{q_1}\mu_{q_2}}{q_1 q_2} \ , \\ 
    Z_3(\bq_1,\bq_2, \bq_3,a) & =  K_3(\bq_1,\bq_2, a)+ f \mu^2_k G_3(\bq_1,\bq_2,\bq_3) \nonumber \\
    & + f \mu_k k \Bigg[\frac{\mu_{q_1}}{q_1}K_2(\bq_2,\bq_3) + \frac{\mu_{q_{23}}}{q_{23}}G_2(\bq_2,\bq_3) \Bigg(b_1 + f\mu_k k \frac{\mu_{q_1}}{q_1} \Bigg)  \nonumber \\
    & + b_1 f \mu_k k \frac{\mu_{q_2}\mu_{q_3}}{q_2 q_3} + 2 \ \text{cyclic} \Bigg] \nonumber \\
    & + f^3 \mu_k^3 k^3 \frac{\mu_{q_1}\mu_{q_2}\mu_{q_3}}{q_1 q_2 q_3}
\end{align}

where we have used the notation $\mu_k \equiv \hat{\bk} \cdot \hat{\mathbf{z}}$, $\mu_i \equiv \hat{\bq}_i \cdot \hat{\mathbf{z}}$, $\mu_{ij} \equiv \hat{\bq}_{ij} \cdot \hat{\mathbf{z}}$, and $\bq_{12\ldots n} \equiv \bq_1 +\bq_2 + \ldots + \bq_n$.

\section{Galaxy Power Spectrum and Bispectrum in redshift space}
\label{sec:theoretical_ps_bisp}

The galaxy 1-loop power spectrum in redshift space is given by 

\begin{align}
\label{ps_1loop}
    P^{\text{1-loop}}_{g,s}(k; a) & = Z_1(\bk;a)^2P_L(k; a)  \nonumber \\ 
    & + 2 \int\frac{d^3\bq}{(2\pi)^3}\big[Z_2(\bk-\bq,\bq; a)^2\big]P_L(q; a)P_L(|\bk-\bq|; a) \nonumber \\
    & + 6 Z_1(k;a)P_L(k; a)\int\frac{d^3\bq}{(2\pi)^3}Z_3(\bk,\bq,-\bq; a)P_L(q;a) \nonumber \\
    & + P_{ct}(\bk; a) + P_{\epsilon}(\bk; a) 
\end{align}
where $P_L(k; a)$ denotes the linear power spectrum, defined as
\begin{equation}
    \langle \delta_{g,s}^{(1)}(\bk, a)\delta_g^{(1)}(\bk',a)\rangle = (2\pi)^3\delta_D(\bk+\bk')P_L(k; a),
\end{equation}
and $P_{ct}(\bk,a)$ and $P_{\epsilon}(\bk, a)$ are respectively the counterterms and the stochastic terms, that account for the impact of small-scale physics on large scale modes. They are given by \cite{DAmico:2022ukl, DAmico:2022gki}
\begin{align}
    P_{ct}(\bk, a) & = 2P_L(k)Z_1(\bk) \bigg(\frac{k}{k_M} \bigg)^2 \big(c_{ct} + c_{cr,1}\mu_k^2 + c_{cr,2}\mu_k^4 \big) , \\
    P_{\epsilon}(\bk, a) & = \frac{1}{\bar{n}_g} \Bigg(c_{\epsilon, 0} + \bigg(\frac{k}{k_M} \bigg)^2 c_{\epsilon, 1} + f \mu_k^2 \bigg(\frac{k}{k_M} \bigg)^2 c_{\epsilon, 2} \Bigg)\,,
\end{align}
where the pivot scale $k_M$ is of the order of the typical comoving scale of halos, $\bar{n}_g$ is the mean number density of galaxies and we have used the notation $\mu_k \equiv \hat{\bk} \cdot \hat{\mathbf{z}}$, being $\hat{z}$ the line of sight.

The angle-averaged multipoles of the redshift-space power spectrum are defined as

\begin{equation}
    P_{\ell}(k,z) = \frac{2\ell+1}{2}\int_{-1}^{1}d\mu_k P^{\text{1-loop}}_g(k,\mu_k;z)\mathcal{P}_\ell(\mu_k)\,,
    \label{eq:PSmult}
\end{equation}
where $\mathcal{P}_\ell(\mu_k)$ are the Legendre polynomials of order $\ell$. In this work we will consider the monopole $(\ell=0)$, the quadrupole $(\ell=2)$ and the hexadecapole $(\ell=4)$ of the galaxy power spectrum.

The tree-level bispectrum is given by 
\begin{align}
    B(\bq_1, \bq_2, \bq_3) & = 2 Z_1(\bq_1)Z_1(\bq_2)Z_2(\bq_1,\bq_2)P_L(q_1)P_L(q_2) + 2 \ \text{perms.} \nonumber \\
    & + B_{\epsilon}(\bq_1,\bq_2,\bq_3) ,
\end{align}
where the shot noise contribution is \cite{DAmico:2022ukl, DAmico:2022osl},

\begin{equation}
    B_{\epsilon}(\bq_1,\bq_2,\bq_3) = \frac{ c_{\epsilon,3}}{\bar{n}_g^2}  + \frac{1}{\bar{n}_g}\sum_{i=1}^3\big(b_1 + f (\hat{\bq}_i\cdot\hat{z})^2 \big) \left( c_{\epsilon,4} + c_{\epsilon, 0} f (\hat{\bq}_i\cdot\hat{z})^2 \right) P_L(q_i) \,. 
\end{equation}

The bispectrum monopole and quadrupole are defined as
\begin{align}
    B_0(q_1, q_2, q_3) &= \frac{1}{4\pi} \int_{-1}^1 d\mu_1 \int_0^{2\pi} d\phi B(q_1, q_2, q_3, \mu_1, \mu_2(\mu_1,\phi)) \ , \\
    B_2(q_1, q_2, q_3) & = \frac{5}{4\pi} \int_{-1}^1 d\mu_1 \int_0^{2\pi} d\phi \mathcal{P}_2(\mu_3(\mu_1,\phi))B(q_1, q_2, q_3, \mu_1, \mu_2(\mu_1, \phi)) \ .
\end{align}
where we follow the conventions adopted in \cite{DAmico:2022osl, DAmico:2022ukl}, with 
\begin{align}
    \mu_2 &\equiv \mu_1 \hat{q}_1\cdot \hat{q}_2 - \sqrt{1-\mu_1^2} \sqrt{1-(\hat{q}_1\cdot \hat{q}_2)^2} \cos{\phi} \ , \\
    \mu_3 & \equiv -q_3^{-1} \big(q_1 \mu_1 + q_2 \mu_2(\mu_1, \phi)\big).
\end{align}
Further details on the implementation can be found in these references.

%%%%%%%%%%%%%%%%%%%%%%%%%%%%%%%

\section{Impact of $A_s$ Prior Choice}
\label{app:As_comparison}

\begin{figure}[H]
  \centering
  \begin{subfigure}{0.49\textwidth}
    \includegraphics[width=\linewidth]{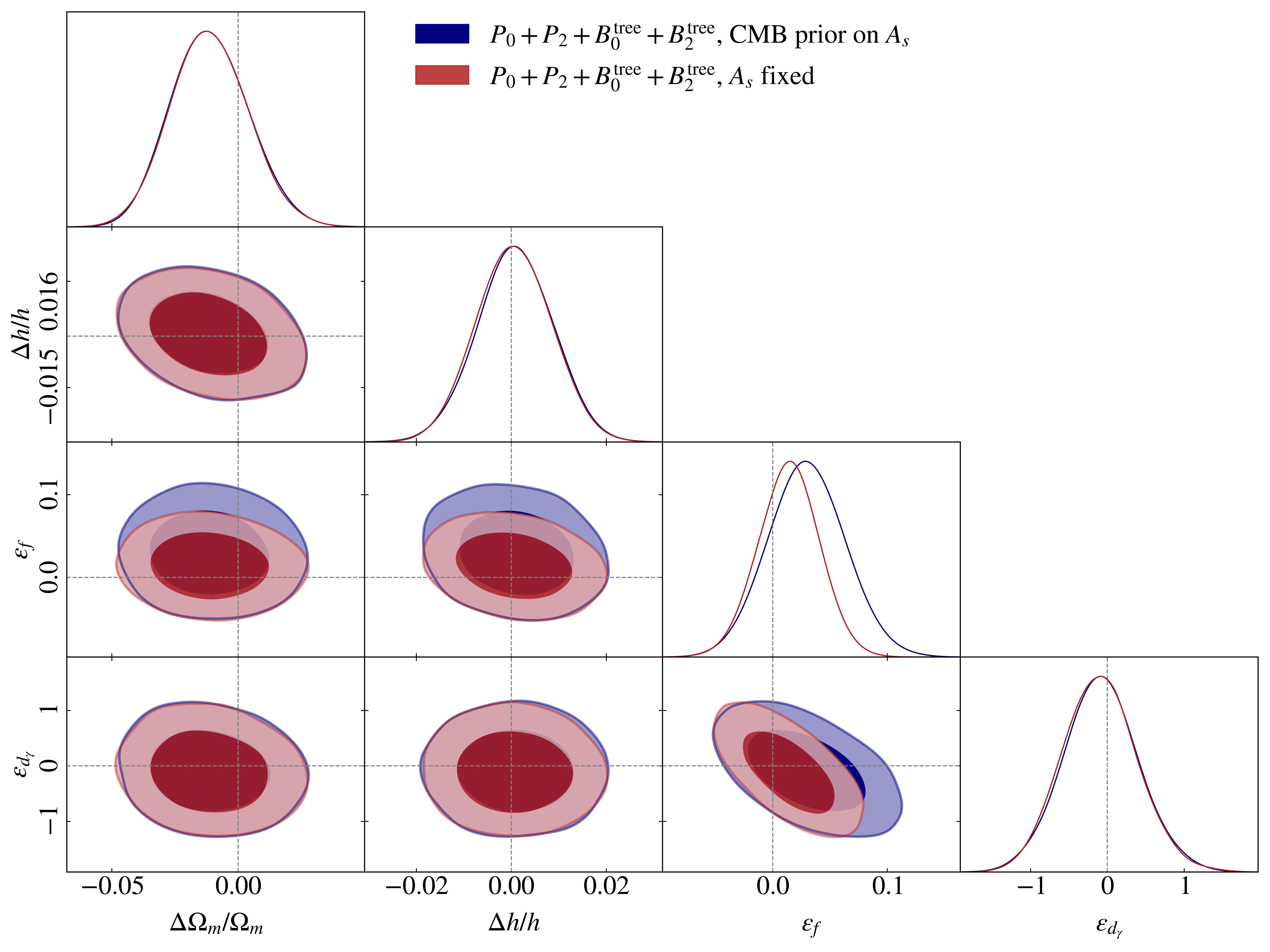}
    \caption{PT Challenge single-box}
    \label{fig:comparison_As_single_volume}
  \end{subfigure}
  \hfill
  \begin{subfigure}{0.49\textwidth}
    \includegraphics[width=\linewidth]{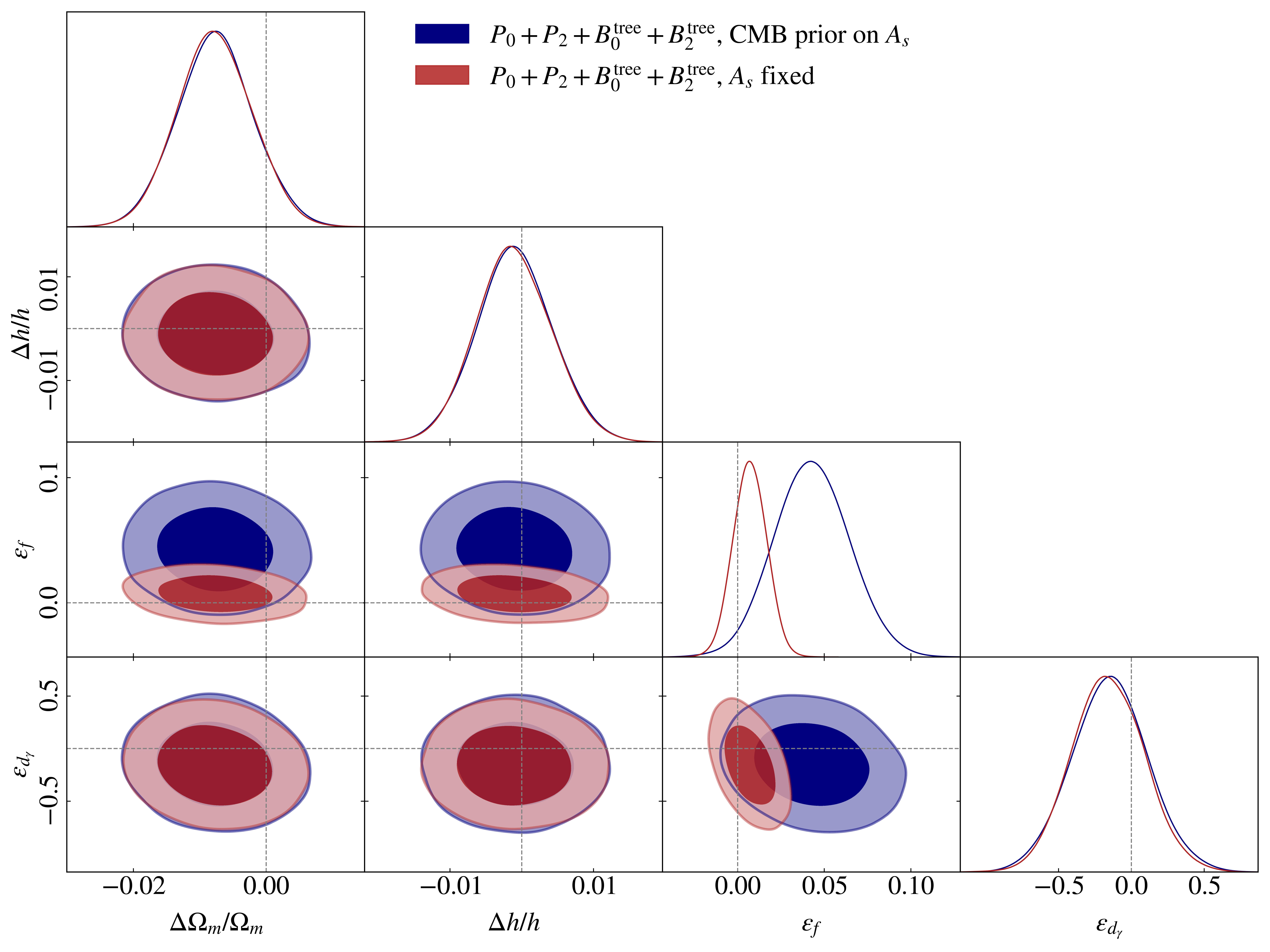}
    \caption{PT Challenge full-volume} 
    \label{fig:comparison_As_th_cov}
  \end{subfigure}
 \caption{Posterior distributions for the PT Challenge simulations, comparing results obtained by fixing $A_s$ to the true value of the simulations and by imposing a $3\sigma$ prior from \textit{Planck}. On the left we show the results using the volume of a single simulation box, while on the right we consider the full volume covariance. We adopt $k_{\text{max},B_0} = 0.07\, \ihMpc$ and $k_{\text{max},B_2} = 0.04\, \ihMpc$. Dashed lines, centered at 0, represent the true values of the simulations.}
  \label{fig:comparison_As}
\end{figure}

To assess how strongly our findings depend on fixing the amplitude parameter $A_s$ to the true value used in the simulations, we repeat the analysis, this time imposing on $A_s$ a 3$\sigma$ Gaussian prior derived from \textit{Planck} \cite{Planck2018}, following the approach of \cite{Tsedrik:2022cri, Carrilho:2022mon}. 
Figure~\ref{fig:comparison_As} shows the posterior distributions for both the single-box (Fig.~\ref{fig:comparison_As_single_volume}) and the full-volume (10 boxes, Fig.~\ref{fig:comparison_As_th_cov}) of the PT Challenge simulations.

As expected, changing the prior on $A_s$ affects the parameter $\varepsilon_f$, owing to their degeneracy, discussed in Sect.~\ref{sec:degeneracies}.
In the single-volume case, applying the \textit{Planck} prior results in a slightly wider posterior for $\varepsilon_f$, but the corresponding shift in its central value remains well below $1\sigma$. 
When considering the entire simulation volume, the \textit{Planck} prior yields a broader posterior (still consistent at the $1\sigma$ level with that obtained for fixed $A_s$) and a larger shift in the central value of $\varepsilon_f$, which, however, still lies within $2\sigma$ of the true value.

For the parameter $\varepsilon_{d_{\gamma}}$, the posterior distributions remain unchanged when the priors on $A_s$ are varied in either volume setup. This indicates that the parameter is only weakly affected by degeneracies, thereby reinforcing its robustness as a probe of departures from $\Lambda$CDM.

%%%%%%%%%%%%%%%%%%%%%%%%%%%%%%%%%%%%%%%%%%

\section{Theoretical covariance}
\label{app:theoretical_cov}

For the analysis of the PT Challenge simulations we employ a theoretical model for the covariance matrix of the power spectrum and the bispectrum, given the limited number of realizations. We briefly review here the derivation and the final formula adopted in the analysis.

\noindent The estimator for the redshift space power spectrum is defined as
\begin{equation}
    \hat{P}_s(\bk) = \frac{1}{V V_s} \int_{\bq \in \bk}d^3\bq \,\ \delta_s(\bq)\delta_s(-\bq)\,,
    \label{eq:PS_est}
\end{equation}
where the shell volume $V_s$ is defined such that eq.~\ref{eq:PS_est} is an unbiased estimator in the thin-shell approximation, $V_s \simeq 4 \pi k^2\delta k$. The estimator for the power spectrum $\ell$-multipole is then given by
\begin{equation}
    \hat{P}_\ell(k)= \frac{2\ell + 1}{2}\int_{-1}^{1}d\mu \,\,\mathcal{L}_l(\mu)\hat{P}_s(\bk)\,.
\end{equation}
One can easily estimate the gaussian covariance for such estimator, obtaining
\begin{equation}
    {\rm Cov}(P_\ell, P_{\ell'})(k, k') = (2\pi)^3\frac{(2\ell + 1)(2\ell'+1)}{2 V V_s} \delta^{\rm K}_{k,k'}\int_{-1}^1 d\mu \,\, \mathcal{L}_{\ell}(\mu)\mathcal{L}_{\ell'}(\mu) P_s^2(\bk)\,,
\end{equation}
where $\delta^{\rm K}$ indicates a Kronecker delta. Notice that the redshift space power spectrum $P_s(\bk)$ in this expression is the linear one, including the constant shot-noise term
\begin{equation}
    P_s(\bk) = (b_1 + f \mu^2)^2 P_L(k) + \frac{1}{\bar{n}}\,.
\end{equation}

For the bispectrum, we consider the following unbiased estimators for the redshift space multipoles~\cite{Scoccimarro:2015bla}
\begin{equation}
    \hat{B}_{\ell}(k_1, k_2, k_3) = \frac{(2\ell +1)}{V V_{123}}\int_{\bq_1\in k_1}d^3\bq_1\int_{\bq_2\in k_2}d^3\bq_2\int_{\bq_3\in k_3}d^3\bq_3\delta_D(\bq_{123})\delta_s(\bq_1)\delta_s(\bq_2)\delta_s(\bq_3)\mathcal{L}_\ell(\mu_1)\,,
\end{equation}
where the integrals over $\bq_i$'s are performed over the shell width $[k_i - \Delta k/2, k_i + \Delta k/2]$, $\mu_1 = \hat{\bq_1}\cdot \hat{\mathbf{n}}$ and $V_{123}$ is the triangular bin volume given by
\begin{equation}
    V_{123} = \int_{k_1}d^3\bq_1\int_{k_1}d^3\bq_2\int_{k_1}d^3\bq_3\,\,\delta_D(\bq_{123}) \simeq 8\pi^2 k_1k_2k_3 \Delta k^3\,.
\end{equation}
Given this estimator, we can compute the covariance among two different triangles $T$ and $T'$ as 
\begin{equation}
    \text{Cov}(T, T') = \langle\hat{B}_\ell(k_1, k_2, k_3)\hat{B}_{\ell'}(k_1', k_2', k_3')\rangle - \langle\hat{B}_\ell(k_1, k_2, k_3)\rangle\langle\hat{B}_{\ell'}(k_1', k_2', k_3')\rangle\,,
\end{equation}
which eventually results in
\begin{equation}
    \text{Cov}(T, T') = \text{Cov}(T, T')^{PPP} + \text{Cov}(T, T')^{BB} + \text{Cov}(T, T')^{PT} + \text{Cov}(T, T')^{6pt}\,.
\end{equation}
The `$PPP$` term denotes the `gaussian` contribution, expressed as the product of three power spectra. The `$BB$` term corresponds to the contribution involving the product of two bispectra, `$PT$` includes the product of a power spectrum and a trispectrum, and `$6pt$` represents the contribution arising from the connected six-point function. We will only include the gaussian term, as the others  have been shown to give negligible  contributions in the range of scales considered in our analysis \cite{Sefusatti:2006pa, Song_2015}. Moreover, for the same reason, in the PT challenge we neglect the contribution to the covariance arising from the cross-correlation between the power spectrum and the bispectrum, as the analysis is restricted to large scales where this term is negligible. For the analysis of BOSS data, we will make use of the numerical covariance matrix estimated from 2048 Patchy mocks~\cite{Kitaura_2016}.
It is given by
\begin{equation}
\begin{split}
    \text{Cov}(T, T')^{PPP} = \frac{(2\ell +1)(2\ell' +1)}{V^2 V_{123}^2}&\int_{\bq_1}\int_{\bq_2}\int_{\bq_3}\int_{\bq_1'}\int_{\bq_2'}\int_{\bq_3'}\delta_D(\bq_{123})\delta_D(\bq_{123}')\mathcal{L}_\ell(\mu_1)\mathcal{L}_\ell(\mu_1') \times\\
    &\times\Big(\langle\delta_s(\bq_1)\delta_s(\bq_1')\rangle\langle\delta_s(\bq_2)\delta_s(\bq_2')\rangle\langle\delta_s(\bq_3)\delta_s(\bq_3')\rangle + \text{5 perms.}\Big)\,.
\end{split}
\label{eq:covPPP}
\end{equation}
We first integrate over $\bq_1'$, $\bq_2'$ and $\bq_3'$, and exploit the definition $V = (2\pi)^3\delta_D(0)$ to simplify the expression
\begin{equation}
\begin{split}
    \text{Cov}(T, T')^{PPP} = (2\pi)^6\frac{(2\ell +1)(2\ell' +1)}{V V_{123}^2}&\int_{\bq_1}\int_{\bq_2}\int_{\bq_3}\delta_D(\bq_{123})\Big[c_1\mathcal{L}_\ell(\mu_1)\mathcal{L}_\ell(-\mu_1) + c_2\mathcal{L}_\ell(\mu_1)\mathcal{L}_\ell(-\mu_2) + c_3\mathcal{L}_\ell(\mu_1)\mathcal{L}_\ell(-\mu_3)\Big] \\
    & \times P_s(\bq_1)P_s(\bq_2)P_s(\bq_3)\,.
\end{split}
\end{equation}
To further simplify the expression for the bispectrum covariance, we expand the power spectra in multipoles using
\begin{equation}
    P_s(\bk) = \sum_{\ell}\mathcal{L}_{\ell}(\mu)P_\ell(k)\,,
\end{equation}
such that
\begin{equation}
\begin{split}
    \text{Cov}(T, T')^{PPP} = (2\pi)^6\frac{(2\ell +1)(2\ell' +1)}{V V_{123}^2}&\int_{\bq_1}\int_{\bq_2}\int_{\bq_3}\delta_D(\bq_{123})\Big[c_1\mathcal{L}_\ell(\mu_1)\mathcal{L}_\ell(-\mu_1) + c_2\mathcal{L}_\ell(\mu_1)\mathcal{L}_\ell(-\mu_2) + c_3\mathcal{L}_\ell(\mu_1)\mathcal{L}_\ell(-\mu_3)\Big] \\
    & \times \mathcal{L}_{\ell_1}(\mu_1)\mathcal{L}_{\ell_2}(\mu_2)\mathcal{L}_{\ell_3}(\mu_3)\sum_{\ell_1, \ell_2, \ell_3}P_{\ell_1}(q_1)P_{\ell_2}(q_2)P_{\ell_3}(q_3)\,,
\end{split}
\end{equation}
which can be rewritten in the compact form (assuming thin-enough bins)
\begin{equation}
    \text{Cov}(T, T')^{PPP} = (2\pi)^6\frac{(2\ell +1)(2\ell' +1)}{V V_{123}}\sum_{\ell_1, \ell_2, \ell_3} F_{\ell_1\ell_2\ell_3}^{\ell\ell'}(k_1, k_2, k_3)P_{\ell_1}(k_1)P_{\ell_2}(k_2)P_{\ell_3}(k_3)\,,
    \label{eq:cov_bs}
\end{equation}
with
\begin{equation}
    F_{\ell_1\ell_2\ell_3}^{\ell\ell'}(k_1, k_2, k_3) = \frac{1}{4\pi}\int^{2\pi}_0 d\xi \int_{-1}^{1}d\mu_1\Big[c_1 \mathcal{L}_{\ell}(\mu_1)\mathcal{L}_{\ell'}(-\mu_1) + c_2 \mathcal{L}_{\ell}(\mu_1)\mathcal{L}_{\ell'}(-\mu_2) + c_3 \mathcal{L}_{\ell}(\mu_1)\mathcal{L}_{\ell'}(-\mu_3)\Big]\mathcal{L}_{\ell_1}(\mu_1)\mathcal{L}_{\ell_2}(\mu_2)\mathcal{L}_{\ell_3}(\mu_3)\,.
\end{equation}
The coefficients $c_1$, $c_2$ and $c_3$ are symmetry coefficients such that
\begin{equation}
    \begin{split}
        &\{c_1, c_2, c_3\} = \{2,2,2\} \quad \text{if} \quad k_1 = k_2 = k_3\,,\\
        &\{c_1, c_2, c_3\} = \{2,0,0\} \quad \text{if} \quad k_2 = k_3\,,\\
        &\{c_1, c_2, c_3\} = \{1,1,0\} \quad \text{if} \quad k_1 = k_2\,,\\
        &\{c_1, c_2, c_3\} = \{1,0,0\} \quad \text{if} \quad k_1 \neq k_2 \neq k_3 \,.
    \end{split}
\end{equation}
Equation~\eqref{eq:cov_bs}, already found in~\cite{Rizzo:2022lmh}, is the expression we will use to compute the covariance of the bispectrum multipoles. The number density and volume inserted in these expressions are the ones of the PT challenge, reported in section~\ref{subsec:PTch}. For the bias parameters we adopt the ones estimated in appendix~\ref{app:full_posteriors}.

%%%%%%%%%%%%%%%%%%%%%%%%%%%%%%%%%%%%%%%%

\section{Complete parameter space constraints}
\label{app:full_posteriors}
\begin{figure}[H]
    \centering
    \includegraphics[width=1.0\linewidth]{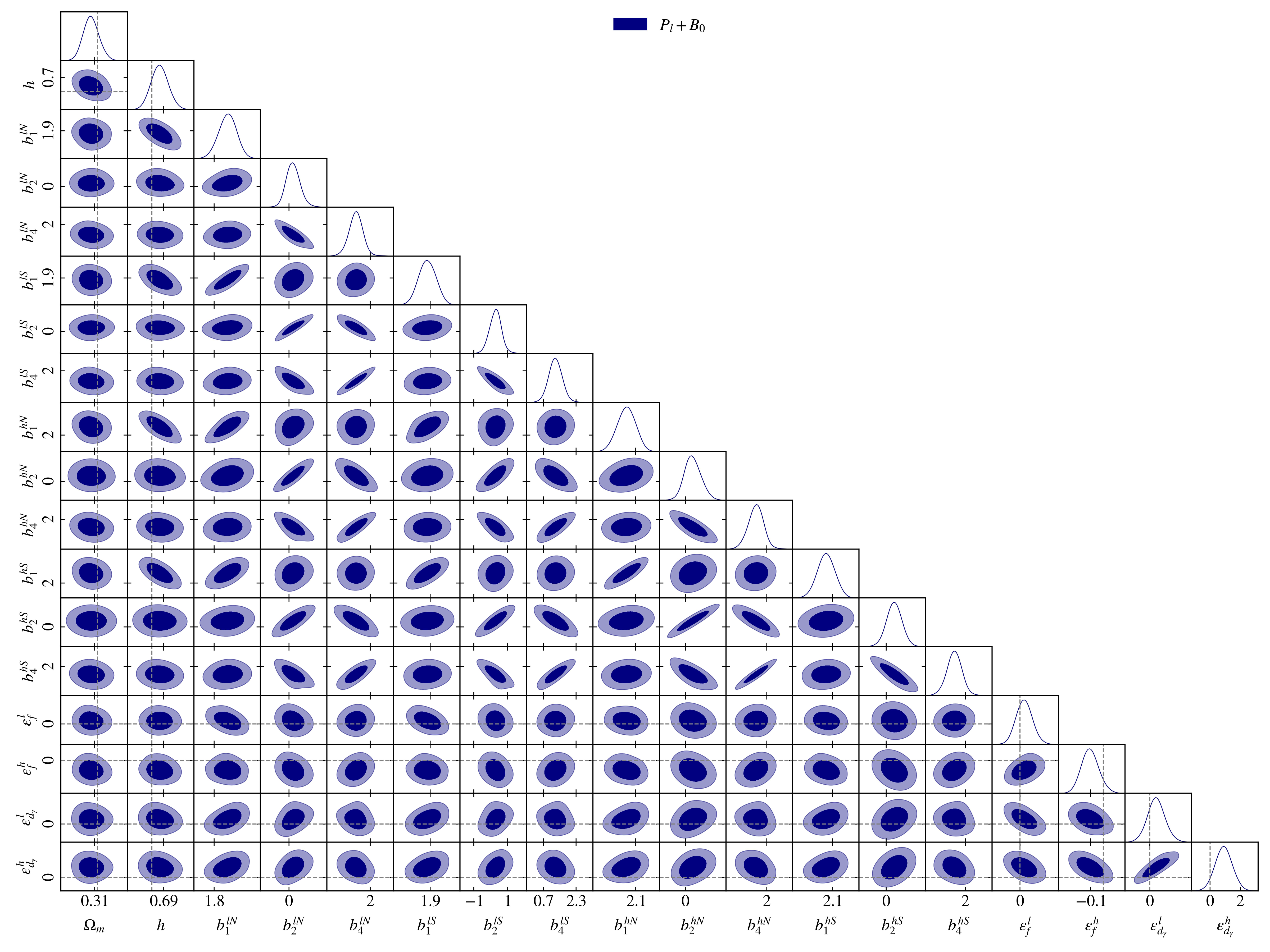}
    \caption{Full posterior distributions from the analysis of BOSS power spectrum multipoles $P_l$ and bispectrum monopole $B_0$, with $n_s$ and $A_s$ fixed to the \textit{Planck} \cite{Planck2018} best-fits. Dashed lines represent the \textit{Planck} best-fit values.} 
    \label{fig:full_posteriors_BOSS}
\end{figure}

\begin{figure}[H]
    \centering
    \includegraphics[width=0.75\linewidth]{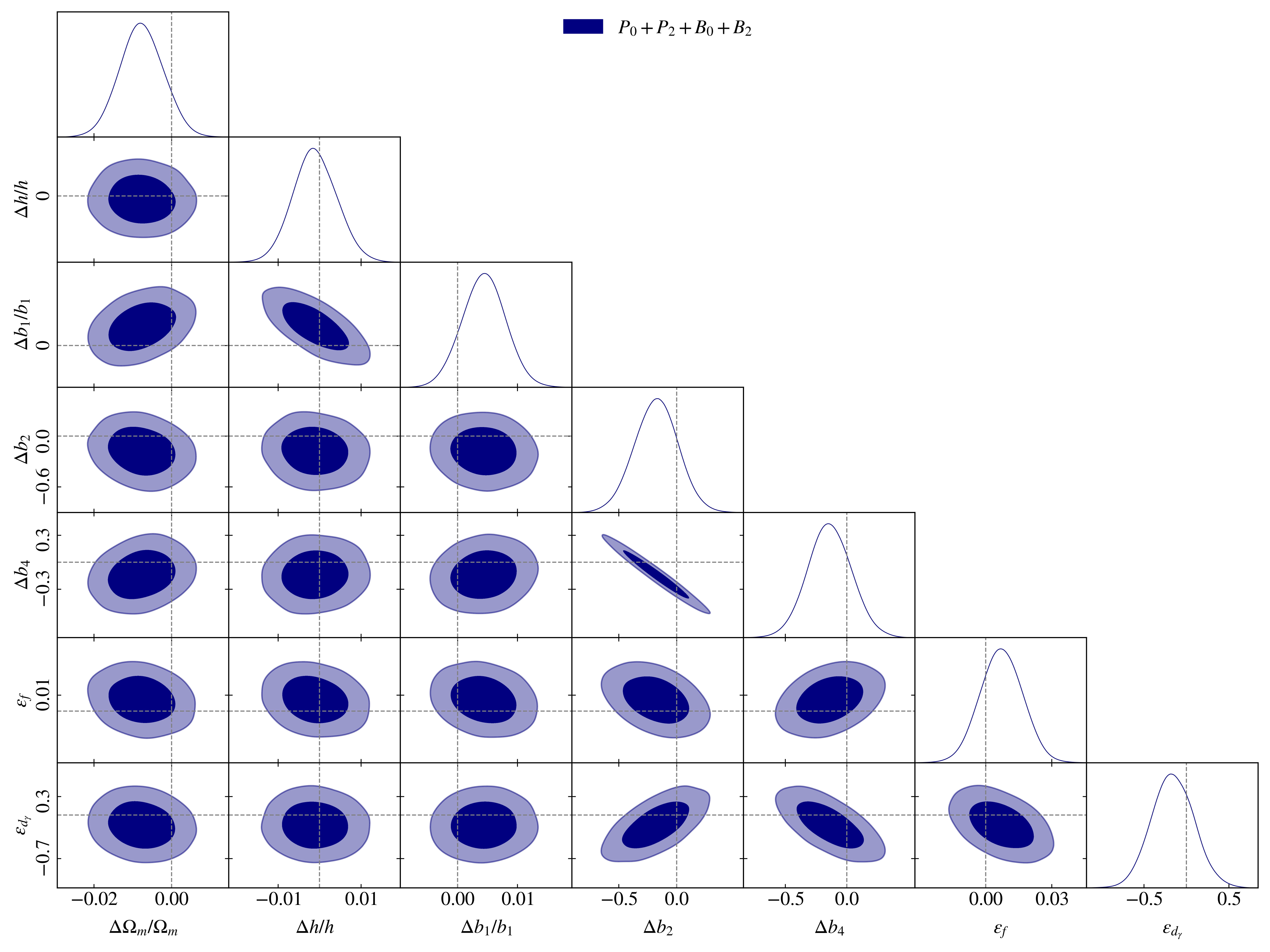}
    \caption{Full posterior distributions for the joint analysis of power spectrum and bispectrum multipoles of the PT Challenge simulations, with $n_s$ and $A_s$ fixed to the true values of the simulations. We adopt $k_{\text{max},B_0} = 0.07 \ihMpc$ and $k_{\text{max},B_2} = 0.04 \ihMpc$. Dashed lines represent the true values of the simulations for the cosmological and bootstrap parameters, and the inferred values for the bias ones (see Appendix \ref{app:full_posteriors}).} 
    \label{fig:full_posteriors_PT_challenge}
\end{figure}

\titlespacing*{\section}{0pt}{10pt}{5pt} 
For completeness, we display the full triangle plots obtained fitting BOSS and PT Challenge power spectrum and bispectrum multipoles, respectively in Fig.~\ref{fig:full_posteriors_BOSS} and Fig.~\ref{fig:full_posteriors_PT_challenge}.  The posterior means with 68\% confidence intervals are reported in Tab.~\ref{tab:bias_BOSS} and Tab.~\ref{tab:bias_PT_challenge}.

\setlength{\intextsep}{6pt}  

\begin{table}[H]
    \centering
    \setlength{\tabcolsep}{12pt}
   \renewcommand{\arraystretch}{1.2} 
    \begin{tabular}{|c|c|c|c|c|}
       \hline
        $\textbf{Dataset}$ & $b_1$ & $b_2$  & $b_4$ \\
        \hline
        LOWZ NGC & $1.877 \pm 0.051$  & $0.25^{+0.38}_{-0.43}$ & $1.32 \pm 0.34$ \\
        LOWZ SGC & $1.884 \pm 0.052$ & $0.29 \pm 0.38$ & $1.30 \pm 0.34$ \\
        CMASS NGC & $2.055 \pm 0.045$ & $0.45^{+0.45}_{-0.53}$ & $1.46 \pm 0.38$  \\
        CMASS SGC & $2.069 \pm 0.045$ & $0.50 \pm 0.42$ & $1.44 \pm 0.38$ \\
        \hline
    \end{tabular}
    \caption{Constraints on bias parameters $b_1$, $b_2$ and $b_4$ from the BOSS LOWZ and CMASS Northern and Southern galactic caps (NGC and SGC) datasets, with $n_s$ and $A_s$ fixed to \textit{Planck} \cite{Planck2018} best-fits. Reported values correspond to posterior means with 68\% confidence intervals.}
    \label{tab:bias_BOSS}
\end{table}

\begin{table}[H]
    \centering
    \setlength{\tabcolsep}{12pt}
    \renewcommand{\arraystretch}{1.2} 
    \begin{tabular}{|c|c|c|c|}
        \hline
        $\textbf{Dataset}$ & $\Delta b_1/b_1$ & $\Delta b_2$  & $\Delta b_4$ \\
        \hline
        $P_0+P_2+B_0+B_2$ & $0.0044 \pm 0.0037$ & $-0.18\pm 0.19$ & $-0.14 \pm 0.18 $\\
        \hline
    \end{tabular}
    \caption{Constraints on bias parameters $b_1$, $b_2$ and $b_4$ from the PT Challenge simulations, with $n_s$ and $A_s$ fixed to the true values. Reported values correspond to posterior means with 68\% confidence intervals. Here, $\Delta b_i = b_i - \bar{b}_i$ denotes the deviation of bias parameter $b_i$ from its value inferred as discussed in Appendix \ref{app:full_posteriors}.}
    \label{tab:bias_PT_challenge}
\end{table}

The estimated true values for the bias parameters reported in Table~\ref{tab:bias_PT_challenge} were obtained using the peak-background split (PBS) approach~\cite{Desjacques:2016bnm}, see also~\cite{Kaiser:1984sw, Bardeen:1985tr}. In this framework, (local) halo biases are interpreted as the response of the halo abundance to long-wavelength density perturbations. To compute such response, the key ingredient is the halo mass function, which describes the number distribution of halos given a specific halo mass. On top of that, the second key ingredient is the halo occupation distribution (HOD): this describes how galaxies populate halos of different masses. The procedure to estimate the halo mass function and the HOD for the PT Challenge simulations is described in~\cite{Nishimichi:2020tvu}, and we adopted their function to estimate the bias parameters. Given these two ingredients, one can estimate the bias parameters as
\begin{equation}
    b_{\mathcal{O}}(z) = \frac{\int dM \frac{d n_{h}}{dM}(M, z) N_g(M,z) b_{\mathcal{O}}(M,z)}{\int dM \frac{d n_{h}}{dM}(M, z) N_g(M,z)}\,,
    \label{eq:bias_PBS}
\end{equation}
where $d n_{h}(M, z)/dM$ is the halo mass function, $N_g(M,z)$ is HOD function. The mass dependent biases $b_{\mathcal{O}}(M, z)$ can be obtained assuming universality of the halo mass function. One of the best empirical functions for the linear bias $b_1$ is~\cite{Tinker:2010my}
\begin{equation}
    b_1(M) = 1 - A \frac{\nu(M)^a}{\nu(M)^a + \delta_c^a} + B\nu(M)^b + C\nu(M)^c
    \label{eq:b1_Tinker}
\end{equation}
with $\delta_c = 1.686$ being the critical overdensity for spherical collapse, $\nu$ is the peak height defined as
\begin{equation}
    \nu(M) = \frac{\delta_c}{\sigma(M)}\,,
\end{equation}
with $\sigma(M)$ defined as the linear matter variance defined on the Lagrangian scale of the halo $R = (3 M / 4 \pi \bar{\rho}_{m})^{1/3}$ filtered with a top-hat window function
\begin{equation}
    \sigma^2(M, z) \equiv \int d^{3}\bq P_{\rm lin}(k,z) W^2(k, R(M))\,.
\end{equation}
The parameters in eq.~\ref{eq:b1_Tinker} are obtained by fitting to different simulations
\begin{equation}
    \begin{aligned}
        A & = 1.0 + 0.24\,y\,e^{-(4/y)^4} \,,
& \quad
a & = 0.44 y - 0.88\,, \\
B & = 0.183 \,,
& \quad
b & = 1.5 \,,\\
C &= 0.019 + 0.107\,y + 0.19\,e^{-(4/y)^4}\,,\quad  &  c &= 2.4\,,\nonumber
    \end{aligned}
\end{equation}
where $y = \log{200}$.
Following the notation of~\cite{Lazeyras:2017hxw}, we can write the galaxy density perturbation field at second perturbative order as
\begin{equation}
\delta_g^{(2)}(\bx) = b_1 \delta^{(2)}(\bx) + \frac{b_{\delta^2}}{2}\big[\delta^{(1)2} - \langle\delta^{(1)2}\rangle\big] + b_{K^2}\big[K^{2}(\bx) -\langle K^2\rangle\big]\,,
    \label{eq:2nd_delta}
\end{equation}
where the tidal operator is defined as
\begin{equation}
    K_{ij}(\bx) = \left[\frac{\partial_i \partial_j}{\nabla^2} - \frac{1}{3}\delta_{ij}^{\rm K}\right] \delta(\bx)\,.
\end{equation}
The PBS approach cannot predict non-local bias parameters, for which more empirical relations were found: we will make use of the relation between the linear second order bias $b_{\delta^2}$ and $b_1$ found in~\cite{Lazeyras:2015lgp}
\begin{equation}
    b_{\delta^2}(M) = 0.412 - 2.143\,b_1(M) + 0.929\,b_1^2(M) + 0.008\,b_1^3(M)\,.
    \label{eq:b2_Laz}
\end{equation}
PBS does not provide for an analytical expression for the non-local bias parameters~\cite{Desjacques:2016bnm}. One can see that the local Lagrangian approximation (Lagrangian local-in-matter-density, LLIMD) constrains the shape of the tidal bias to be
\begin{equation}
    b_{K^2}^{\rm LLIMD}(M) = - \frac{2}{7}\Big(b_1(M) - 1\Big)\,,
\end{equation}
which has been shown to be a poor fit to actual data, see~\cite{Lazeyras:2017hxw, Eggemeier:2021cam, Akitsu:2024lyt, Ivanov:2025qie}. Methods based on the excursion-set approach predict a slightly modified version of the LLIMD relation, yielding~\cite{Sheth:2012fc}
\begin{equation}
    b_{K^2}^{\rm Ex.\,set}(M) = 0.524 - 0.547 b_1(M) + 0.046 b_1^2(M)\,,
\end{equation}
which seems to agree better with data from simulations~\cite{Eggemeier:2021cam}. Here we will adopt a refined version of this relation based on the findings of~\cite{Akitsu:2024lyt}
\begin{equation}
    b_{K^2}(M) = 0.932 - 1.297\,b_1(M) + 0.345\,b_1^2(M) - 0.034\,b_1^3(M)\,.
    \label{eq:bK2_Kazu}
\end{equation}
After inserting eqs.~\ref{eq:b2_Laz} and \ref{eq:bK2_Kazu} into eq.~\ref{eq:bias_PBS} one can easily calculate the expected values for the bias parameters $b_1(z)$, $b_{\delta^2}(z)$ and  $b_{K^2}(z)$ at a given redshift $z$ and compare with the best-fit results.
Let us stress that the second order biases shown in eqs.~\ref{eq:b2_Laz} and \ref{eq:bK2_Kazu} are equivalent to the bias parameters adopted in this work. The mapping between the two bases is explicitly given by
\begin{equation}
    b_2(z) = \frac{7}{2}b_{K^2}(z) + b_1(z)\,,\qquad b_4(z) = \frac{1}{2}b_{\delta^2}(z) - \frac{17}{6}b_{K^2}(z)\,.
    \label{eq:b24K}
\end{equation}
The values computed with this procedure are compared with those obtained by fitting to the PT Challenge data, and the comparison is shown in figures~\ref{fig:full_posteriors_PT_challenge} and table~\ref{tab:bias_PT_challenge}, which results in a good agreement.

%%%%%%%%%%%%%%%%%%%%%

%\section{Testing the pipeline}
%\label{app:test}

%To test and validate the pipeline, we first perform analyses on the PT Challenge simulations within a fixed $\Lambda$CDM cosmology. Marginalized posteriors are shown in Fig.~\ref{fig:LCDM_test}.

%\begin{figure}[t]
%    \centering
    %\includegraphics[width=0.75\linewidth]{img/PT_challenge_LCDM_test.jpg}
    %\caption{Full posterior distributions for the joint analysis of power spectrum and bispectrum multipoles of the PT Challenge simulations, with $n_s$ and $A_s$ fixed to the true values of the simulations. We adopt $k_{\text{max},B_0} = 0.07 \ihMpc$ and $k_{\text{max},B_2} = 0.04 \ihMpc$. Dashed lines represent the true values of the simulations.} 
    %\label{fig:LCDM_test}
%\end{figure}

\bibliography{ref}

@article{Scoccimarro:2015bla,
    author = "Scoccimarro, Roman",
    title = "{Fast Estimators for Redshift-Space Clustering}",
    eprint = "1506.02729",
    archivePrefix = "arXiv",
    primaryClass = "astro-ph.CO",
    doi = "10.1103/PhysRevD.92.083532",
    journal = "Phys. Rev. D",
    volume = "92",
    number = "8",
    pages = "083532",
    year = "2015"
}

@article{Simon:2022lde,
    author = "Simon, Th{\'e}o and Zhang, Pierre and Poulin, Vivian and Smith, Tristan L.",
    title = "{Consistency of effective field theory analyses of the BOSS power spectrum}",
    eprint = "2208.05929",
    archivePrefix = "arXiv",
    primaryClass = "astro-ph.CO",
    doi = "10.1103/PhysRevD.107.123530",
    journal = "Phys. Rev. D",
    volume = "107",
    number = "12",
    pages = "123530",
    year = "2023"
}

@article{Beyond-2pt:2024mqz,
    author = "Krause, Elisabeth and others",
    collaboration = "Beyond-2pt",
    title = "{A Parameter-Masked Mock Data Challenge for Beyond-Two-Point Galaxy Clustering Statistics}",
    eprint = "2405.02252",
    archivePrefix = "arXiv",
    primaryClass = "astro-ph.CO",
    doi = "10.3847/1538-4357/ad781d",
    journal = "Astrophys. J.",
    volume = "990",
    number = "2",
    pages = "99",
    year = "2025"
}

@article{Marinucci:2024bdq,
    author = "Marinucci, Marco and others",
    title = "{The constraining power of the marked power spectrum: an analytical study}",
    eprint = "2411.14377",
    archivePrefix = "arXiv",
    primaryClass = "astro-ph.CO",
    doi = "10.1088/1475-7516/2025/09/036",
    journal = "JCAP",
    volume = "09",
    pages = "036",
    year = "2025"
}

@article{Peron:2024xaw,
    author = "Peron, Matteo and Jung, Gabriel and Liguori, Michele and Pietroni, Massimo",
    title = "{Constraining primordial non-Gaussianity from large scale structure with the wavelet scattering transform}",
    eprint = "2403.17657",
    archivePrefix = "arXiv",
    primaryClass = "astro-ph.CO",
    doi = "10.1088/1475-7516/2024/07/021",
    journal = "JCAP",
    volume = "07",
    pages = "021",
    year = "2024"
}

@article{Amendola:2022vte,
    author = "Amendola, Luca and Pietroni, Massimo and Quartin, Miguel",
    title = "{Fisher matrix for the one-loop galaxy power spectrum: measuring expansion and growth rates without assuming a cosmological model}",
    eprint = "2205.00569",
    archivePrefix = "arXiv",
    primaryClass = "astro-ph.CO",
    doi = "10.1088/1475-7516/2022/11/023",
    journal = "JCAP",
    volume = "11",
    pages = "023",
    year = "2022"
}

@article{Fujita:2020xtd,
    author = "Fujita, Tomohiro and Vlah, Zvonimir",
    title = "{Perturbative description of biased tracers using consistency relations of LSS}",
    eprint = "2003.10114",
    archivePrefix = "arXiv",
    primaryClass = "astro-ph.CO",
    reportNumber = "CERN-TH-2020-037",
    doi = "10.1088/1475-7516/2020/10/059",
    journal = "JCAP",
    volume = "10",
    pages = "059",
    year = "2020"
}

@article{Lazeyras:2015lgp,
    author = "Lazeyras, Titouan and Wagner, Christian and Baldauf, Tobias and Schmidt, Fabian",
    title = "{Precision measurement of the local bias of dark matter halos}",
    eprint = "1511.01096",
    archivePrefix = "arXiv",
    primaryClass = "astro-ph.CO",
    doi = "10.1088/1475-7516/2016/02/018",
    journal = "JCAP",
    volume = "02",
    pages = "018",
    year = "2016"
}

@article{Tinker:2010my,
    author = "Tinker, Jeremy L. and Robertson, Brant E. and Kravtsov, Andrey V. and Klypin, Anatoly and Warren, Michael S. and Yepes, Gustavo and Gottlober, Stefan",
    title = "{The Large Scale Bias of Dark Matter Halos: Numerical Calibration and Model Tests}",
    eprint = "1001.3162",
    archivePrefix = "arXiv",
    primaryClass = "astro-ph.CO",
    doi = "10.1088/0004-637X/724/2/878",
    journal = "Astrophys. J.",
    volume = "724",
    pages = "878--886",
    year = "2010"
}

@article{Bernardeau:2001qr,
    author = "Bernardeau, F. and Colombi, S. and Gaztanaga, E. and Scoccimarro, R.",
    title = "{Large scale structure of the universe and cosmological perturbation theory}",
    eprint = "astro-ph/0112551",
    archivePrefix = "arXiv",
    reportNumber = "SACLAY-T01-142",
    doi = "10.1016/S0370-1573(02)00135-7",
    journal = "Phys. Rept.",
    volume = "367",
    pages = "1--248",
    year = "2002"
}

@article{Desjacques:2016bnm,
    author = "Desjacques, Vincent and Jeong, Donghui and Schmidt, Fabian",
    title = "{Large-Scale Galaxy Bias}",
    eprint = "1611.09787",
    archivePrefix = "arXiv",
    primaryClass = "astro-ph.CO",
    doi = "10.1016/j.physrep.2017.12.002",
    journal = "Phys. Rept.",
    volume = "733",
    pages = "1--193",
    year = "2018"
}

@article{Wadekar:2020hax,
    author = "Wadekar, Digvijay and Ivanov, Mikhail M. and Scoccimarro, Roman",
    title = "{Cosmological constraints from BOSS with analytic covariance matrices}",
    eprint = "2009.00622",
    archivePrefix = "arXiv",
    primaryClass = "astro-ph.CO",
    reportNumber = "INR-TH-2020-036",
    doi = "10.1103/PhysRevD.102.123521",
    journal = "Phys. Rev. D",
    volume = "102",
    pages = "123521",
    year = "2020"
}

@article{Eggemeier:2021cam,
    author = "Eggemeier, Alexander and Scoccimarro, Rom{\'a}n and Smith, Robert E. and Crocce, Martin and Pezzotta, Andrea and S{\'a}nchez, Ariel G.",
    title = "{Testing one-loop galaxy bias: Joint analysis of power spectrum and bispectrum}",
    eprint = "2102.06902",
    archivePrefix = "arXiv",
    primaryClass = "astro-ph.CO",
    doi = "10.1103/PhysRevD.103.123550",
    journal = "Phys. Rev. D",
    volume = "103",
    number = "12",
    pages = "123550",
    year = "2021"
}

@misc{Akitsu:2024lyt,
    author = "Akitsu, Kazuyuki",
    title = "{Mapping the galaxy-halo connection to the galaxy bias: implication to the HOD-informed prior}",
    eprint = "2410.08998",
    archivePrefix = "arXiv",
    primaryClass = "astro-ph.CO",
    month = "10",
    year = "2024"
}

@misc{Nguyen:2024yth,
    author = "Nguyen, Nhat-Minh and Schmidt, Fabian and Tucci, Beatriz and Reinecke, Martin and Kosti\'c, Andrija",
    title = "{How much information can be extracted from galaxy clustering at the field level?}",
    eprint = "2403.03220",
    archivePrefix = "arXiv",
    primaryClass = "astro-ph.CO",
    reportNumber = "LCTP-24-05",
    month = "3",
    year = "2024"
}

@article{Planck2018,
    author = "Aghanim, N. and others",
    collaboration = "Planck",
    title = "{Planck 2018 results. VI. Cosmological parameters}",
    eprint = "1807.06209",
    archivePrefix = "arXiv",
    primaryClass = "astro-ph.CO",
    doi = "10.1051/0004-6361/201833910",
    journal = "Astron. Astrophys.",
    volume = "641",
    pages = "A6",
    year = "2020",
    note = "[Erratum: Astron.Astrophys. 652, C4 (2021)]"
}

@article{Desjacques:2018pfv,
    author = "Desjacques, Vincent and Jeong, Donghui and Schmidt, Fabian",
    title = "{The Galaxy Power Spectrum and Bispectrum in Redshift Space}",
    eprint = "1806.04015",
    archivePrefix = "arXiv",
    primaryClass = "astro-ph.CO",
    doi = "10.1088/1475-7516/2018/12/035",
    journal = "JCAP",
    volume = "12",
    pages = "035",
    year = "2018"
}

@article{Euclid:2024yrr,
    author = "Mellier, Y. and others",
    collaboration = "Euclid",
    title = "{Euclid. I. Overview of the Euclid mission}",
    eprint = "2405.13491",
    archivePrefix = "arXiv",
    primaryClass = "astro-ph.CO",
    doi = "10.1051/0004-6361/202450810",
    journal = "Astron. Astrophys.",
    volume = "697",
    pages = "A1",
    year = "2025"
}

@misc{DESI:2016fyo,
    author = "Aghamousa, A. and others",
    collaboration = "DESI",
    title = "{The DESI Experiment Part I: Science,Targeting, and Survey Design}",
    eprint = "1611.00036",
    archivePrefix = "arXiv",
    primaryClass = "astro-ph.IM",
    reportNumber = "FERMILAB-PUB-16-517-AE",
    month = "10",
    year = "2016"
}

@article{Sheth:2012fc,
    author = "Sheth, Ravi K. and Chan, Kwan Chuen and Scoccimarro, Rom{\'a}n",
    title = "{Nonlocal Lagrangian bias}",
    eprint = "1207.7117",
    archivePrefix = "arXiv",
    primaryClass = "astro-ph.CO",
    doi = "10.1103/PhysRevD.87.083002",
    journal = "Phys. Rev. D",
    volume = "87",
    number = "8",
    pages = "083002",
    year = "2013"
}

@article{Kaiser:1984sw,
    author = "Kaiser, Nick",
    title = "{On the Spatial correlations of Abell clusters}",
    doi = "10.1086/184341",
    journal = "Astrophys. J. Lett.",
    volume = "284",
    pages = "L9--L12",
    year = "1984"
}

@article{Bardeen:1985tr,
    author = "Bardeen, James M. and Bond, J. R. and Kaiser, Nick and Szalay, A. S.",
    title = "{The Statistics of Peaks of Gaussian Random Fields}",
    reportNumber = "FERMILAB-PUB-85-148-A, NSF-ITP-85-93",
    doi = "10.1086/164143",
    journal = "Astrophys. J.",
    volume = "304",
    pages = "15--61",
    year = "1986"
}

@article{Ivanov:2023qzb,
    author = "Ivanov, Mikhail M. and Philcox, Oliver H. E. and Cabass, Giovanni and Nishimichi, Takahiro and Simonovi\'c, Marko and Zaldarriaga, Matias",
    title = "{Cosmology with the galaxy bispectrum multipoles: Optimal estimation and application to BOSS data}",
    eprint = "2302.04414",
    archivePrefix = "arXiv",
    primaryClass = "astro-ph.CO",
    reportNumber = "YITP-23-13, CERN-TH-2023-022",
    doi = "10.1103/PhysRevD.107.083515",
    journal = "Phys. Rev. D",
    volume = "107",
    number = "8",
    pages = "083515",
    year = "2023"
}

@article{Philcox:2022frc,
    author = "Philcox, Oliver H. E. and Ivanov, Mikhail M. and Cabass, Giovanni and Simonovi\'c, Marko and Zaldarriaga, Matias and Nishimichi, Takahiro",
    title = "{Cosmology with the redshift-space galaxy bispectrum monopole at one-loop order}",
    eprint = "2206.02800",
    archivePrefix = "arXiv",
    primaryClass = "astro-ph.CO",
    reportNumber = "CERN-TH-2022-092, YITP-22-60",
    doi = "10.1103/PhysRevD.106.043530",
    journal = "Phys. Rev. D",
    volume = "106",
    number = "4",
    pages = "043530",
    year = "2022"
}

@article{Philcox:2021kcw,
    author = "Philcox, Oliver H. E. and Ivanov, Mikhail M.",
    title = "{BOSS DR12 full-shape cosmology: {\ensuremath{\Lambda}}CDM constraints from the large-scale galaxy power spectrum and bispectrum monopole}",
    eprint = "2112.04515",
    archivePrefix = "arXiv",
    primaryClass = "astro-ph.CO",
    doi = "10.1103/PhysRevD.105.043517",
    journal = "Phys. Rev. D",
    volume = "105",
    number = "4",
    pages = "043517",
    year = "2022"
}

@article{Piga:2022mge,
    author = "Piga, Lorenzo and Marinucci, Marco and D'Amico, Guido and Pietroni, Massimo and Vernizzi, Filippo and Wright, Bill S.",
    title = "{Constraints on modified gravity from the BOSS galaxy survey}",
    eprint = "2211.12523",
    archivePrefix = "arXiv",
    primaryClass = "astro-ph.CO",
    doi = "10.1088/1475-7516/2023/04/038",
    journal = "JCAP",
    volume = "04",
    pages = "038",
    year = "2023"
}

@article{DAmico:2022osl,
    author = "D'Amico, Guido and Donath, Yaniv and Lewandowski, Matthew and Senatore, Leonardo and Zhang, Pierre",
    title = "{The BOSS bispectrum analysis at one loop from the Effective Field Theory of Large-Scale Structure}",
    eprint = "2206.08327",
    archivePrefix = "arXiv",
    primaryClass = "astro-ph.CO",
    reportNumber = "NUHEP-TH/22-05",
    doi = "10.1088/1475-7516/2024/05/059",
    journal = "JCAP",
    volume = "05",
    pages = "059",
    year = "2024"
}

@article{DAmico:2022ukl,
    author = "D'Amico, Guido and Donath, Yaniv and Lewandowski, Matthew and Senatore, Leonardo and Zhang, Pierre",
    title = "{The one-loop bispectrum of galaxies in redshift space from the Effective Field Theory of Large-Scale Structure}",
    eprint = "2211.17130",
    archivePrefix = "arXiv",
    primaryClass = "astro-ph.CO",
    doi = "10.1088/1475-7516/2024/07/041",
    journal = "JCAP",
    volume = "07",
    pages = "041",
    year = "2024"
}

@article{DAmico:2022gki,
    author = "D'Amico, Guido and Lewandowski, Matthew and Senatore, Leonardo and Zhang, Pierre",
    title = "{Limits on primordial non-Gaussianities from BOSS galaxy-clustering data}",
    eprint = "2201.11518",
    archivePrefix = "arXiv",
    primaryClass = "astro-ph.CO",
    doi = "10.1103/PhysRevD.111.063514",
    journal = "Phys. Rev. D",
    volume = "111",
    number = "6",
    pages = "063514",
    year = "2025"
}

@article{DAmico:2019fhj,
    author = "D'Amico, Guido and Gleyzes, J\'er\^ome and Kokron, Nickolas and Markovic, Katarina and Senatore, Leonardo and Zhang, Pierre and Beutler, Florian and Gil-Mar\'\i{}n, H\'ector",
    title = "{The Cosmological Analysis of the SDSS/BOSS data from the Effective Field Theory of Large-Scale Structure}",
    eprint = "1909.05271",
    archivePrefix = "arXiv",
    primaryClass = "astro-ph.CO",
    doi = "10.1088/1475-7516/2020/05/005",
    journal = "JCAP",
    volume = "05",
    pages = "005",
    year = "2020"
}

@article{DAmico:2020kxu,
    author = "D'Amico, Guido and Senatore, Leonardo and Zhang, Pierre",
    title = "{Limits on $w$CDM from the EFTofLSS with the PyBird code}",
    eprint = "2003.07956",
    archivePrefix = "arXiv",
    primaryClass = "astro-ph.CO",
    doi = "10.1088/1475-7516/2021/01/006",
    journal = "JCAP",
    volume = "01",
    pages = "006",
    year = "2021"
}

@article{Marinucci:2024add,
    author = "Marinucci, Marco and Pardede, Kevin and Pietroni, Massimo",
    title = "{Bootstrapping Lagrangian perturbation theory for the large scale structure}",
    eprint = "2405.08413",
    archivePrefix = "arXiv",
    primaryClass = "astro-ph.CO",
    doi = "10.1088/1475-7516/2024/10/051",
    journal = "JCAP",
    volume = "10",
    pages = "051",
    year = "2024"
}

@article{Amendola:2023awr,
    author = "Amendola, Luca and Marinucci, Marco and Pietroni, Massimo and Quartin, Miguel",
    title = "{Improving precision and accuracy in cosmology with model-independent spectrum and bispectrum}",
    eprint = "2307.02117",
    archivePrefix = "arXiv",
    primaryClass = "astro-ph.CO",
    doi = "10.1088/1475-7516/2024/01/001",
    journal = "JCAP",
    volume = "01",
    pages = "001",
    year = "2024"
}

@misc{Ansari:2025nsf,
    author = "Ansari, Arhum and Banerjee, Arka and Jain, Sachin and Lalsodagar, Sahil",
    title = "{Bootstrapping LSS perturbation theory beyond third order}",
    eprint = "2504.01078",
    archivePrefix = "arXiv",
    primaryClass = "astro-ph.CO",
    month = "4",
    year = "2025"
}

@article{Chudaykin:2025aux,
    author = "Chudaykin, Anton and Ivanov, Mikhail M. and Philcox, Oliver H. E.",
    title = "{Reanalyzing DESI DR1. I. {\ensuremath{\Lambda}}CDM constraints from the power spectrum and bispectrum}",
    eprint = "2507.13433",
    archivePrefix = "arXiv",
    primaryClass = "astro-ph.CO",
    reportNumber = "MIT-CPT/5890",
    doi = "10.1103/qsnt-dppc",
    journal = "Phys. Rev. D",
    volume = "113",
    number = "6",
    pages = "063502",
    year = "2026"
}

@misc{Chudaykin:2026nls,
    author = "Chudaykin, Anton and Ivanov, Mikhail M. and Philcox, Oliver H. E.",
    title = "{Reanalyzing DESI DR1: 5. Cosmological Constraints with Simulation-Based Priors}",
    eprint = "2602.18554",
    archivePrefix = "arXiv",
    primaryClass = "astro-ph.CO",
    reportNumber = "MIT-CTP/6007",
    month = "2",
    year = "2026"
}

@misc{Chudaykin:2025vdh,
    author = "Chudaykin, Anton and Ivanov, Mikhail M. and Philcox, Oliver H. E.",
    title = "{Reanalyzing DESI DR1: 3. Constraints on Inflation from Galaxy Power Spectra {\&} Bispectra}",
    eprint = "2512.04266",
    archivePrefix = "arXiv",
    primaryClass = "astro-ph.CO",
    reportNumber = "MIT-CTP/5971",
    month = "12",
    year = "2025"
}

@misc{Chudaykin:2025lww,
    author = "Chudaykin, Anton and Ivanov, Mikhail M. and Philcox, Oliver H. E.",
    title = "{Reanalyzing DESI DR1: 2. Constraints on Dark Energy, Spatial Curvature, and Neutrino Masses}",
    eprint = "2511.20757",
    archivePrefix = "arXiv",
    primaryClass = "astro-ph.CO",
    reportNumber = "MIT-CTP/5960",
    month = "11",
    year = "2025"
}

@article{Peron:2025lgh,
    author = "Peron, Matteo and Nishimichi, Takahiro and Pietroni, Massimo and Taruya, Atsushi",
    title = "{Renormalized perturbation theory at field-level: the LSS bootstrap in GridSPT}",
    eprint = "2506.07105",
    archivePrefix = "arXiv",
    primaryClass = "astro-ph.CO",
    reportNumber = "YITP-25-88",
    doi = "10.1088/1475-7516/2025/10/098",
    journal = "JCAP",
    volume = "10",
    pages = "098",
    year = "2025"
}

@article{Nishimichi:2020tvu,
    author = "Nishimichi, Takahiro and D'Amico, Guido and Ivanov, Mikhail M. and Senatore, Leonardo and Simonovi\'c, Marko and Takada, Masahiro and Zaldarriaga, Matias and Zhang, Pierre",
    title = "{Blinded challenge for precision cosmology with large-scale structure: results from effective field theory for the redshift-space galaxy power spectrum}",
    eprint = "2003.08277",
    archivePrefix = "arXiv",
    primaryClass = "astro-ph.CO",
    reportNumber = "YITP-20-25, INR-TH-2020-009, CERN-TH-2020-040, IPMU20-0025",
    doi = "10.1103/PhysRevD.102.123541",
    journal = "Phys. Rev. D",
    volume = "102",
    number = "12",
    pages = "123541",
    year = "2020"
}

@article{Ivanov:2021kcd,
    author = "Ivanov, Mikhail M. and Philcox, Oliver H. E. and Nishimichi, Takahiro and Simonovi\'c, Marko and Takada, Masahiro and Zaldarriaga, Matias",
    title = "{Precision analysis of the redshift-space galaxy bispectrum}",
    eprint = "2110.10161",
    archivePrefix = "arXiv",
    primaryClass = "astro-ph.CO",
    reportNumber = "YITP-21-120, CERN-TH-2021-155",
    doi = "10.1103/PhysRevD.105.063512",
    journal = "Phys. Rev. D",
    volume = "105",
    number = "6",
    pages = "063512",
    year = "2022"
}

@article{Sefusatti:2006pa,
    author = "Sefusatti, Emiliano and Crocce, Martin and Pueblas, Sebastian and Scoccimarro, Roman",
    title = "{Cosmology and the Bispectrum}",
    eprint = "astro-ph/0604505",
    archivePrefix = "arXiv",
    reportNumber = "FERMILAB-PUB-06-083-A",
    doi = "10.1103/PhysRevD.74.023522",
    journal = "Phys. Rev. D",
    volume = "74",
    pages = "023522",
    year = "2006"
}

@article{Taule:2024bot,
    author = "Taule, Petter and Marinucci, Marco and Biselli, Giorgia and Pietroni, Massimo and Vernizzi, Filippo",
    title = "{Constraints on dark energy and modified gravity from the BOSS Full-Shape and DESI BAO data}",
    eprint = "2409.08971",
    archivePrefix = "arXiv",
    primaryClass = "astro-ph.CO",
    doi = "10.1088/1475-7516/2025/03/036",
    journal = "JCAP",
    volume = "03",
    pages = "036",
    year = "2025"
}

@article{DAmico:2021rdb,
    author = "D'Amico, Guido and Marinucci, Marco and Pietroni, Massimo and Vernizzi, Filippo",
    title = "{The large scale structure bootstrap: perturbation theory and bias expansion from symmetries}",
    eprint = "2109.09573",
    archivePrefix = "arXiv",
    primaryClass = "astro-ph.CO",
    doi = "10.1088/1475-7516/2021/10/069",
    journal = "JCAP",
    volume = "10",
    pages = "069",
    year = "2021"
}

@article{Carrasco:2012cv,
    author = "Carrasco, John Joseph M. and Hertzberg, Mark P. and Senatore, Leonardo",
    title = "{The Effective Field Theory of Cosmological Large Scale Structures}",
    eprint = "1206.2926",
    archivePrefix = "arXiv",
    primaryClass = "astro-ph.CO",
    doi = "10.1007/JHEP09(2012)082",
    journal = "JHEP",
    volume = "09",
    pages = "082",
    year = "2012"
}

@article{Baumann:2010tm,
    author = "Baumann, Daniel and Nicolis, Alberto and Senatore, Leonardo and Zaldarriaga, Matias",
    title = "{Cosmological Non-Linearities as an Effective Fluid}",
    eprint = "1004.2488",
    archivePrefix = "arXiv",
    primaryClass = "astro-ph.CO",
    doi = "10.1088/1475-7516/2012/07/051",
    journal = "JCAP",
    volume = "07",
    pages = "051",
    year = "2012"
}

@article{DESI:2025zgx,
    author = "Abdul Karim, M. and others",
    collaboration = "DESI",
    title = "{DESI DR2 results. II. Measurements of baryon acoustic oscillations and cosmological constraints}",
    eprint = "2503.14738",
    archivePrefix = "arXiv",
    primaryClass = "astro-ph.CO",
    reportNumber = "FERMILAB-PUB-25-0169-PPD",
    doi = "10.1103/tr6y-kpc6",
    journal = "Phys. Rev. D",
    volume = "112",
    number = "8",
    pages = "083515",
    year = "2025"
}

@article{DESI:2024mwx,
    author = "Adame, A. G. and others",
    collaboration = "DESI",
    title = "{DESI 2024 VI: cosmological constraints from the measurements of baryon acoustic oscillations}",
    eprint = "2404.03002",
    archivePrefix = "arXiv",
    primaryClass = "astro-ph.CO",
    reportNumber = "FERMILAB-PUB-24-0154-PPD",
    doi = "10.1088/1475-7516/2025/02/021",
    journal = "JCAP",
    volume = "02",
    pages = "021",
    year = "2025"
}

@article{DESI:2024jxi,
    author = "Adame, A. G. and others",
    collaboration = "DESI",
    title = "{DESI 2024 V: Full-Shape galaxy clustering from galaxies and quasars}",
    eprint = "2411.12021",
    archivePrefix = "arXiv",
    primaryClass = "astro-ph.CO",
    reportNumber = "FERMILAB-PUB-24-0847-PPD",
    doi = "10.1088/1475-7516/2025/09/008",
    journal = "JCAP",
    volume = "09",
    pages = "008",
    year = "2025",
    note = "[Erratum: JCAP 02, E02 (2026)]"
}

@article{DESI:2025fii,
    author = "Lodha, K. and others",
    collaboration = "DESI",
    title = "{Extended dark energy analysis using DESI DR2 BAO measurements}",
    eprint = "2503.14743",
    archivePrefix = "arXiv",
    primaryClass = "astro-ph.CO",
    reportNumber = "FERMILAB-PUB-25-0164-PPD",
    doi = "10.1103/w4c6-1r5j",
    journal = "Phys. Rev. D",
    volume = "112",
    number = "8",
    pages = "083511",
    year = "2025"
}

@misc{Ivanov:2025qie,
    author = "Ivanov, Mikhail M.",
    title = "{Simulation-Based Priors without Simulations: an Analytic Perspective on EFT Parameters of Galaxies}",
    eprint = "2503.07270",
    archivePrefix = "arXiv",
    primaryClass = "astro-ph.CO",
    reportNumber = "MIT-CTP/5851",
    month = "3",
    year = "2025"
}

@article{Lazeyras:2017hxw,
    author = "Lazeyras, Titouan and Schmidt, Fabian",
    title = "{Beyond LIMD bias: a measurement of the complete set of third-order halo bias parameters}",
    eprint = "1712.07531",
    archivePrefix = "arXiv",
    primaryClass = "astro-ph.CO",
    doi = "10.1088/1475-7516/2018/09/008",
    journal = "JCAP",
    volume = "09",
    pages = "008",
    year = "2018"
}

@article{Rizzo:2022lmh,
    author = "Rizzo, Federico and Moretti, Chiara and Pardede, Kevin and Eggemeier, Alexander and Oddo, Andrea and Sefusatti, Emiliano and Porciani, Cristiano and Monaco, Pierluigi",
    title = "{The halo bispectrum multipoles in redshift space}",
    eprint = "2204.13628",
    archivePrefix = "arXiv",
    primaryClass = "astro-ph.CO",
    doi = "10.1088/1475-7516/2023/01/031",
    journal = "JCAP",
    volume = "01",
    pages = "031",
    year = "2023"
}

@misc{brinckmann2018montepython3boostedmcmc,
      title={MontePython 3: boosted MCMC sampler and other features}, 
      author={Thejs Brinckmann and Julien Lesgourgues},
      year={2018},
      eprint={1804.07261},
      archivePrefix={arXiv},
      primaryClass={astro-ph.CO},
      url={https://arxiv.org/abs/1804.07261}, 
}

@article{Audren:2012wb,
      author         = "Audren, Benjamin and Lesgourgues, Julien and Benabed,
                        Karim and Prunet, Simon",
      title          = "{Conservative Constraints on Early Cosmology: an
                        illustration of the Monte Python cosmological parameter
                        inference code}",
      journal        = "JCAP",
      volume         = "1302",
      pages          = "001",
      doi            = "10.1088/1475-7516/2013/02/001",
      year           = "2013",
      eprint         = "1210.7183",
      archivePrefix  = "arXiv",
      primaryClass   = "astro-ph.CO",
      reportNumber   = "CERN-PH-TH-2012-290, LAPTH-048-12",
      SLACcitation   = "%%CITATION = ARXIV:1210.7183;%%",
}

@article{Alam_2017,
   title={The clustering of galaxies in the completed SDSS-III Baryon Oscillation Spectroscopic Survey: cosmological analysis of the DR12 galaxy sample},
   volume={470},
   ISSN={1365-2966},
   url={http://dx.doi.org/10.1093/mnras/stx721},
   DOI={10.1093/mnras/stx721},
   number={3},
   journal={Monthly Notices of the Royal Astronomical Society},
   publisher={Oxford University Press (OUP)},
   author={Alam, Shadab and Ata, Metin and Bailey, Stephen and Beutler, Florian and Bizyaev, Dmitry and Blazek, Jonathan A. and Bolton, Adam S. and Brownstein, Joel R. and Burden, Angela and Chuang, Chia-Hsun and Comparat, Johan and Cuesta, Antonio J. and Dawson, Kyle S. and Eisenstein, Daniel J. and Escoffier, Stephanie and Gil-Marín, Héctor and Grieb, Jan Niklas and Hand, Nick and Ho, Shirley and Kinemuchi, Karen and Kirkby, David and Kitaura, Francisco and Malanushenko, Elena and Malanushenko, Viktor and Maraston, Claudia and McBride, Cameron K. and Nichol, Robert C. and Olmstead, Matthew D. and Oravetz, Daniel and Padmanabhan, Nikhil and Palanque-Delabrouille, Nathalie and Pan, Kaike and Pellejero-Ibanez, Marcos and Percival, Will J. and Petitjean, Patrick and Prada, Francisco and Price-Whelan, Adrian M. and Reid, Beth A. and Rodríguez-Torres, Sergio A. and Roe, Natalie A. and Ross, Ashley J. and Ross, Nicholas P. and Rossi, Graziano and Rubiño-Martín, Jose Alberto and Saito, Shun and Salazar-Albornoz, Salvador and Samushia, Lado and Sánchez, Ariel G. and Satpathy, Siddharth and Schlegel, David J. and Schneider, Donald P. and Scóccola, Claudia G. and Seo, Hee-Jong and Sheldon, Erin S. and Simmons, Audrey and Slosar, Anže and Strauss, Michael A. and Swanson, Molly E. C. and Thomas, Daniel and Tinker, Jeremy L. and Tojeiro, Rita and Magaña, Mariana Vargas and Vazquez, Jose Alberto and Verde, Licia and Wake, David A. and Wang, Yuting and Weinberg, David H. and White, Martin and Wood-Vasey, W. Michael and Yèche, Christophe and Zehavi, Idit and Zhai, Zhongxu and Zhao, Gong-Bo},
   year={2017},
   month=mar, pages={2617–2652} }

@misc{reid2015sdssiiibaryonoscillationspectroscopic,
      title={SDSS-III Baryon Oscillation Spectroscopic Survey Data Release 12: galaxy target selection and large scale structure catalogues}, 
      author={Beth Reid and Shirley Ho and Nikhil Padmanabhan and Will J. Percival and Jeremy Tinker and Rita Tojeiro and Martin White and Daniel J. Eisenstein and Claudia Maraston and Ashley J. Ross and Ariel G. Sanchez and David Schlegel and Erin Sheldon and Michael A. Strauss and Daniel Thomas and David Wake and Florian Beutler and Dmitry Bizyaev and Adam S. Bolton and Joel R. Brownstein and Chia-Hsun Chuang and Kyle Dawson and Paul Harding and Francisco-Shu Kitaura and Alexie Leauthaud and Karen Masters and Cameron K. McBride and Surhud More and Matthew D. Olmstead and Daniel Oravetz and Sebastian E. Nuza and Kaike Pan and John Parejko and Janine Pforr and Francisco Prada and Sergio Rodriguez-Torres and Salvador Salazar-Albornoz and Lado Samushia and Donald P. Schneider and Claudia G. Scoccola and Audrey Simmons and Mariana Vargas-Magana},
      year={2015},
      eprint={1509.06529},
      archivePrefix={arXiv},
      primaryClass={astro-ph.CO},
      url={https://arxiv.org/abs/1509.06529}, 
}

@article{Tsedrik:2022cri,
    author = "Tsedrik, Maria and Moretti, Chiara and Carrilho, Pedro and Rizzo, Federico and Pourtsidou, Alkistis",
    title = "{Interacting dark energy from the joint analysis of the power spectrum and bispectrum multipoles with the EFTofLSS}",
    eprint = "2207.13011",
    archivePrefix = "arXiv",
    primaryClass = "astro-ph.CO",
    doi = "10.1093/mnras/stad260",
    journal = "Mon. Not. Roy. Astron. Soc.",
    volume = "520",
    number = "2",
    pages = "2611--2632",
    year = "2023"
}

@article{Carrilho:2022mon,
    author = "Carrilho, Pedro and Moretti, Chiara and Pourtsidou, Alkistis",
    title = "{Cosmology with the EFTofLSS and BOSS: dark energy constraints and a note on priors}",
    eprint = "2207.14784",
    archivePrefix = "arXiv",
    primaryClass = "astro-ph.CO",
    doi = "10.1088/1475-7516/2023/01/028",
    journal = "JCAP",
    volume = "01",
    pages = "028",
    year = "2023"
}

@article{Gelman:1992zz,
    author = "Gelman, Andrew and Rubin, Donald B.",
    title = "{Inference from Iterative Simulation Using Multiple Sequences}",
    doi = "10.1214/ss/1177011136",
    journal = "Statist. Sci.",
    volume = "7",
    pages = "457--472",
    year = "1992"
}

@article{Lewis_2025,
   title={GetDist: a Python package for analysing Monte Carlo samples},
   volume={2025},
   ISSN={1475-7516},
   url={http://dx.doi.org/10.1088/1475-7516/2025/08/025},
   DOI={10.1088/1475-7516/2025/08/025},
   number={08},
   journal={Journal of Cosmology and Astroparticle Physics},
   publisher={IOP Publishing},
   author={Lewis, Antony},
   year={2025},
   month=aug, pages={025} }

@misc{perko2016biasedtracersredshiftspace,
      title={Biased Tracers in Redshift Space in the EFT of Large-Scale Structure}, 
      author={Ashley Perko and Leonardo Senatore and Elise Jennings and Risa H. Wechsler},
      year={2016},
      eprint={1610.09321},
      archivePrefix={arXiv},
      primaryClass={astro-ph.CO},
      url={https://arxiv.org/abs/1610.09321}, 
}

@misc{Bakx:2025pop,
    author = "Bakx, Thomas and Ivanov, Mikhail M. and Philcox, Oliver H. E. and Vlah, Zvonimir",
    title = "{One-Loop Galaxy Bispectrum: Consistent Theory, Efficient Analysis with COBRA, and Implications for Cosmological Parameters}",
    eprint = "2507.22110",
    archivePrefix = "arXiv",
    primaryClass = "astro-ph.CO",
    reportNumber = "MIT-CTP/5891, RBI-ThPhys-2025-23",
    month = "7",
    year = "2025"
}

@article{Song_2015,
   title={Cosmology with anisotropic galaxy clustering from the combination of power spectrum and bispectrum},
   volume={2015},
   ISSN={1475-7516},
   url={http://dx.doi.org/10.1088/1475-7516/2015/08/007},
   DOI={10.1088/1475-7516/2015/08/007},
   number={08},
   journal={Journal of Cosmology and Astroparticle Physics},
   publisher={IOP Publishing},
   author={Song, Yong-Seon and Taruya, Atsushi and Oka, Akira},
   year={2015},
   month=aug, pages={007–007} }

@article{Alcock:1979mp,
    author = "Alcock, C. and Paczynski, B.",
    title = "{An evolution free test for non-zero cosmological constant}",
    doi = "10.1038/281358a0",
    journal = "Nature",
    volume = "281",
    pages = "358--359",
    year = "1979"
}

@article{Kitaura_2016,
   title={The clustering of galaxies in the SDSS-III Baryon Oscillation Spectroscopic Survey: mock galaxy catalogues for the BOSS Final Data Release},
   volume={456},
   ISSN={1365-2966},
   url={http://dx.doi.org/10.1093/mnras/stv2826},
   DOI={10.1093/mnras/stv2826},
   number={4},
   journal={Monthly Notices of the Royal Astronomical Society},
   publisher={Oxford University Press (OUP)},
   author={Kitaura, Francisco-Shu and Rodríguez-Torres, Sergio and Chuang, Chia-Hsun and Zhao, Cheng and Prada, Francisco and Gil-Marín, Héctor and Guo, Hong and Yepes, Gustavo and Klypin, Anatoly and Scóccola, Claudia G. and Tinker, Jeremy and McBride, Cameron and Reid, Beth and Sánchez, Ariel G. and Salazar-Albornoz, Salvador and Grieb, Jan Niklas and Vargas-Magana, Mariana and Cuesta, Antonio J. and Neyrinck, Mark and Beutler, Florian and Comparat, Johan and Percival, Will J. and Ross, Ashley},
   year={2016},
   month=jan, pages={4156–4173} }

@misc{Spezzati:2025zsb,
    author = "Spezzati, Francesco and Marinucci, Marco and Simonovi{\'c}, Marko",
    title = "{Equivalence of the field-level inference and conventional analyses on large scales}",
    eprint = "2507.05378",
    archivePrefix = "arXiv",
    primaryClass = "astro-ph.CO",
    month = "7",
    year = "2025"
}

@misc{Akitsu:2025boy,
    author = "Akitsu, Kazuyuki and Simonovi{\'c}, Marko and Chen, Shi-Fan and Cabass, Giovanni and Zaldarriaga, Matias",
    title = "{Cosmology inference with perturbative forward modeling at the field level: a comparison with joint power spectrum and bispectrum analyses}",
    eprint = "2509.09673",
    archivePrefix = "arXiv",
    primaryClass = "astro-ph.CO",
    reportNumber = "KEK-Cosmo-0386, RBI-ThPhys-2025-34",
    month = "9",
    year = "2025"
}

\end{document}